\documentclass[prc,twoside,superscriptaddress,showkeys,showpacs,twocolumn]{revtex4}
\usepackage[pdftex,colorlinks,citecolor=blue,bookmarks]{hyperref}
\usepackage{graphicx, latexsym, amssymb, amsmath, color, multirow, mathrsfs, CJK, ifpdf}
\usepackage[section]{placeins}
\usepackage{amsmath}
\usepackage{amsfonts}
\usepackage{amssymb}
\usepackage{bm}
\usepackage{graphicx}
\usepackage{color} %
\usepackage{txfonts}
\makeatletter

\newcommand{\Rmnum}[1]{\expandafter\@slowromancap\romannumeral#1@}
\makeatother

\begin{document}


\title{Triaxial shape fluctuations and quasiparticle excitations in heavy nuclei}
\author{Fang-Qi Chen}
\altaffiliation[Present address: ]
                     {State Key Laboratory of Nuclear Physics and Technology, School of Physics, Peking University, Beijing 100871, China}

\author{J. Luis Egido}
\email{j.luis.egido@uam.es}
\affiliation{Departamento de F\'isica Te\'orica, Universidad Aut\'onoma de Madrid, E-28049, Madrid, Spain}

\date{\today}
\begin{abstract}
The deformation parameters $(\beta,\gamma)$ together with the two-quasiparticle excitations are taken into account, for the first time, as coordinates within a symmetry conserving (angular momentum and particle number)  generator coordinate method.  The simultaneous consideration of collective as well as single particle degrees of freedom allows to describe soft and rigid nuclei as  well as the transition region in between.  

We apply the new theory to the study of the spectra and transition probabilities of the 
$^{156-172}$Er isotopes with a Pairing plus Quadrupole residual interaction. Good agreement with the experimental results is obtained for most  of the observables studied and with the same quality for the very soft and the strongly deformed nuclei.

\end{abstract}

\maketitle


\section{Introduction}\label{sect1}

Beyond mean field theories (BMFT) have developed considerably in recent years \cite{BHR.03,NVR.11,JLE.16}. The consideration of fluctuations around the most probable mean field values and the recovery of the symmetries broken in the mean field approach, in particular the angular momentum, have allowed  us to largely extend the traditional domain of the mean field approach (MFA), in particular to  nuclear spectroscopy. 

These developments have taken place along different lines. The most sophisticated theories start with the mean field approach (in general the Hartree-Fock-Bogoliubov (HFB) one) or with the symmetry conserving mean field approach (SCMFA) \cite{MER.01,JLE.16}. Later on symmetry (angular momentum, AM, particle number, PN, and parity) conserving fluctuations in the most relevant degrees of freedom (shape parameters or energy gaps) are considered within the generator coordinate method (GCM). The solution of the associated Schr\"odinger equation (Hill-Wheeler-Griffin \cite{HWG})  provides energy eigenstates and wave functions. In this line there have been calculations with effective interactions Skyrme \cite{PRC_78_024309_2008}, Gogny \cite{PRC_81_064323_2010,EBR.16} and  relativistic \cite{PRC_81_044311_2010} or schematic interactions \cite{GCM2013}.  A second research line has progressed along the Bohr Collective Hamiltonian \cite{RS.80}, namely the so called five dimensional collective Hamiltonian (5DMCH) derived either from the adiabatic time dependent Hartree-Fock or from the GCM in the Gaussian overlap approximation. There have been calculations with 
schematic interactions \cite{BKColl_PPQ_68} and effective ones, Gogny \cite{GGColl_Gogny_83},  Skyrme \cite{BD.90,PR.09} or relativistic \cite{Niksic_Rel_09} among others.

These investigations  are mainly concerned with the collective degrees of freedom, and an enormous effort, theoretically and numerically, has been made. Important phenomena such as shape coexistence and $\beta$ and $\gamma$ bands, among others,  have been successfully described.

On the opposite end,  non-collective models using  quasiparticle states have been used extensively to study ground and side bands  as in the projected shell model (PSM) \cite{PSMreview,YS.16}.  The natural domain of the PSM is the well deformed nuclei. Here only one shape is considered and by the exhaustive use of multi-quasiparticle states a very good agreement with the ground and non-collective side bands is achieved. The PSM, however, in spite of considering many multi-quasiparticle states, does not always describe properly the collective $\beta$ and $\gamma$ bands. 

The above mentioned GCM theories are complex and CPU time consuming,  therefore,  degrees of freedom such as  two (or more)  quasiparticle states (2qp) have been ignored until now.  The role (absence)  of the 2qp states in these calculations has been minimised (justified) with arguments such as  ``we concentrate on collective states",  or,   ``2qp states that appear at $\beta=\gamma=0$ (for example) appear as ground states at  different  $(\beta,\gamma)$ values considered in the fluctuation mesh".  Though these  arguments are qualitatively correct a quantitative analysis of the role played by pure quasiparticle excitations in {\em collective} states has not been performed.  For example, it is well known that the collectivity of the $\beta-$ and $\gamma-$bands changes with the mass number. In order to shed some light on this issue, in this work we consider simultaneously the PN and AM projected (PNAMP) collective shape fluctuations in the $(\beta,\gamma)$ parameters as well as the PNAMP 2qp states built on top of each shape.  A preliminary analysis along these lines has been performed recently in Ref.~\cite{Fang-Qi_2016} for axially symmetric shapes. 
 We apply the new theory to perform a systematic study of the collective excitations in the Er isotopes from N = 88 to N = 104. Since the light isotopes are very soft in the shape parameters $(\beta,\gamma)$ and the heavy ones are strongly deformed, it is expected that the role played by the collective and the single particle degrees of freedom as well as their coupling will be elucidated. It is expected that the inclusion of the two-quasiparticle states will improve the description of the collective states in the Er isotopes with larger neutron numbers.  In our calculations we use the Pairing plus Quadrupole Hamiltonian of Ref.~\cite{Fang-Qi_2016}.  In Sect.~\ref{sect2} we give a presentation of the theoretical framework and some numerical details. The results are shown and discussed in Sect.~\ref{sect3} followed by a summary and the conclusion in Sect.\ref{sect4}.


\section{Theory}\label{sect2}

 As mentioned in the introduction the shape parameters $(\beta,\gamma)$  are used to generate HFB wave functions, $|\Phi_{0}(\beta,\gamma)\rangle$, with different quadrupolar shapes.  For this purpose we solve the  Hartree-Fock-Bogoliubov  equation with constraints on the total quadrupole moments $\hat{Q}_0\equiv \hat{Q}_{20}$ and $\hat{Q}_2\equiv \hat{Q}_{22}$,  and the average particle number. The
 wave function  $|\Phi_{0}(\beta,\gamma)\rangle$ of the energy minimum for given $(\beta,\gamma)$ values is provided by the solution of the variational principle equation
\begin{equation}
\delta\langle\Phi_{0}(\beta,\gamma)|\hat{H}-\lambda_{n}\hat{N}-\lambda_{p}\hat{Z}-\lambda_{q_0}\hat{Q}_{0} -
\lambda_{q_2}\hat{Q}_{2}|\Phi_{0}(\beta,\gamma)\rangle=0,
\label{HFB_Eq}
\end{equation}
with the Lagrange multipliers $\lambda_{n}$, $\lambda_{p}$, $\lambda_{q_0}$  and $\lambda_{q_2}$ determined by the constraining conditions:
\begin{eqnarray}
\langle\Phi_{0}(\beta,\gamma)|\hat{N}|\Phi_{0}(\beta,\gamma)\rangle=N, \nonumber \\
 \langle\Phi_{0}(\beta,\gamma)|\hat{Z}|\Phi_{0}(\beta,\gamma)\rangle=Z,  \nonumber \\
 \langle\Phi_{0}(\beta,\gamma)|\hat{Q}_{0}|\Phi_{0}(\beta,\gamma)\rangle=q_{0}  \nonumber \\
\langle\Phi_{0}(\beta,\gamma)|\hat{Q}_{2}|\Phi_{0}(\beta,\gamma)\rangle=q_{2}.
\label{Lagr}
\end{eqnarray}
The relation between $(\beta,\gamma)$ and $(q_{0},q_{2})$ is given by 
$\beta= \sqrt{20\pi(q_{0}^{2} + 2 q_{2}^{2})}/3r^{2}_{0}A^{5/3}$,
$\gamma = \arctan (\sqrt{2}{q_2}/q_{0})$
with $r_{0}=1.2$ fm and  $A$ the mass number. These equations are solved for each $(\beta,\gamma)$ point of a grid  in the
$(\beta,\gamma)$ plane.

 For each HFB vacuum  $|\Phi_{0}(\beta,\gamma)\rangle$ there is a set of corresponding quasiparticle operators $\alpha_i(\beta,\gamma)$ satisfying
\begin{equation}
\alpha_i(\beta,\gamma)|\Phi_{0}(\beta,\gamma)\rangle =0,   \;\; \forall i.
\end{equation}
The second components of our GCM ansatz are the two-quasiparticle states, defined by 
\begin{equation}
|\Phi_{ij}(\beta,\gamma)\rangle=\alpha^{\dag}_{i}(\beta,\gamma)\alpha^{\dag}_{j}(\beta,\gamma)|\Phi_{0}(\beta,\gamma)\rangle.
\end{equation}
Finally,  the HFB vacua and the two-quasiparticle states are projected onto good angular momentum and particle number. Thus
 the complete ansatz for our wave function has the form
\begin{eqnarray}\label{eqwf}
|\sigma,IM \rangle&=&\int d\beta d\gamma \sum_{K} f^{\sigma,IK}_{0}(\beta,\gamma) \hat{P}^{I}_{MK}\hat{P}^{N}\hat{P}^{Z} |\Phi_{0}(\beta,\gamma)\rangle\nonumber\\
& & +\sum_{ij}\int d\beta d\gamma \sum_{K} f^{\sigma,IK}_{ij}(\beta,\gamma) \hat{P}^{I}_{MK}\hat{P}^{N}\hat{P}^{Z} |\Phi_{ij}(\beta,\gamma)\rangle\nonumber \\
& = & \sum_{\rho} \int d\beta d\gamma \sum_{K} f^{\sigma,IK}_{\rho}(\beta,\gamma) \hat{P}^{I}_{MK}\hat{P}^{N}\hat{P}^{Z} |\Phi_{\rho}(\beta,\gamma)\rangle\nonumber \\
& = & \sum_{\rho} \int d\beta d\gamma \sum_{K} f^{\sigma,IK}_{\rho}(\beta,\gamma)|IMK,N,\beta,\gamma\rangle_\rho,
\end{eqnarray}
where the index $\rho$ runs over the set $\{0, (ij)\}$ and $\sigma$ labels the different states. We have furthermore introduced the notation $|IMK,N,\beta,\gamma\rangle_\rho$ in an obvious way, the label $Z$ has been suppressed for simplicity.
 The projection operators in the above expression are given by \cite{RS.80}:
\begin{equation}
\hat{P}^{I}_{MK}=\frac{2I+1}{8\pi^{2}}\int d\Omega D^{I*}_{MK}(\Omega)\hat{R}(\Omega),
\end{equation}
for the angular momentum projection, and
\begin{equation}
\hat{P}^{N}=\frac{1}{2\pi}\int^{2\pi}_{0} d\phi \exp[-i\phi(\hat{N}-N)],
\end{equation}
for the particle number projection. It has been shown in Refs.~\cite{Doe.98,MER.01,JLE.16} that the particle number 
projection may cause trouble in the case that the exchange terms of the interaction are neglected.  For this
reason in our calculations we will not neglect any term.

\begin{figure*}[htbp]
	\includegraphics[width=1.0\textwidth]{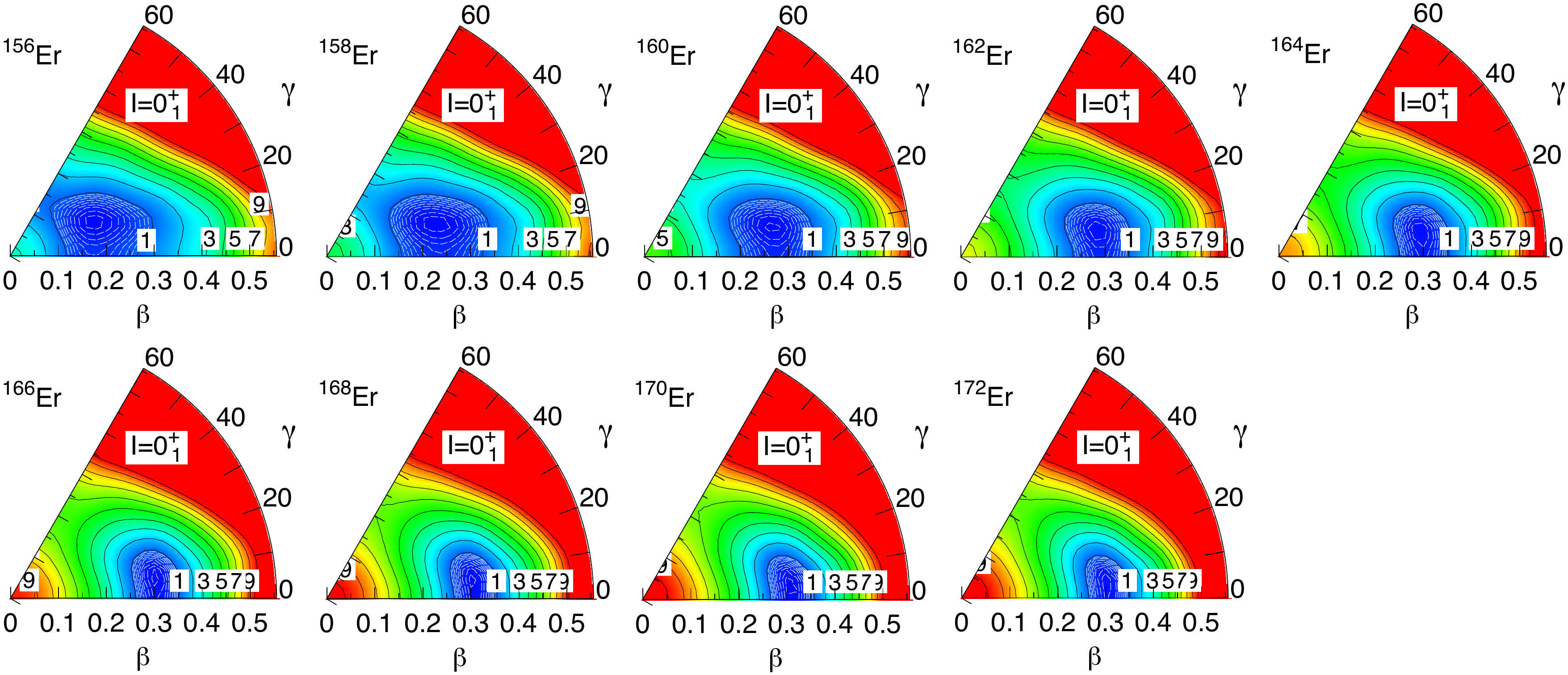}
	\caption{ (Color online) Particle number and angular momentum projected potential energy surfaces  for $^{156-172}$Er for $I=0$ in the $(\beta,\gamma)$ plane. The units for the contours are MeV and for $\gamma$ degrees. The energy origin has been set to zero at the energy minimum in each panel. The black continuous contour lines are 1 MeV apart and the white dashed lines around the minimum are  separated by  0.1 MeV. }
	\label{Erpes0}
\end{figure*}

The variational parameters $f^{\sigma,IK}_{\rho}(\beta,\gamma)$ of Eq.~(\ref{eqwf}) are determined  by minimisation of the energy and this leads to  the Hill-Wheeler-Griffin (HWG) equation  \cite{HWG}
\begin{equation}
\sum_{\rho'\beta'\gamma'K'}\left(\mathcal{H}^{IKK'}_{\rho\rho'}(\beta\gamma, \beta'\gamma')-E^{\sigma I}\mathcal{N}^{IKK'}_{\rho\rho'}(\beta\gamma, \beta'\gamma')\right)f^{\sigma IK'}_{\rho'}(\beta'\gamma')=0
\label{HWG_eq}
\end{equation}
which has to be solved for each value of the angular momentum.  The GCM norm- and energy-overlaps  have been defined as:
\begin{eqnarray}
\mathcal{N}^{IKK'}_{\rho \rho'}(\beta\gamma,\beta'\gamma') &\equiv& _{\rho}\langle IMK,N,\beta,\gamma | IMK',N,\beta',\gamma'\rangle_{\rho'} \nonumber\\
\mathcal{H}^{IKK'}_{\rho \rho'}(\beta\gamma,\beta'\gamma') &\equiv& _{\rho}\langle IMK,N,\beta,\gamma| H |IMK',N,\beta',\gamma'\rangle_{\rho'} .
\label{gcm_overlaps}
\end{eqnarray}
To solve the HWG equations  one first introduces an orthonormal basis defined by the eigenvalues, $n^{\kappa I}$, and eigenvectors, $u^{\kappa I}_{\rho}(\beta,\gamma)$, of the norm overlap:
\begin{equation}
\sum_{\beta'\gamma'\rho'K'}\mathcal{N}^{IKK'}_{\rho \rho'} (\beta\gamma,\beta'\gamma')u^{\kappa IK'}_{\rho'}(\beta'\gamma')=n^{\kappa I}_{}u^{\kappa IK}_{\rho}(\beta,\gamma).
\end{equation}
This orthonormal basis is known as the natural basis and,  for $n^{\kappa I}$ values such that  $n^{\kappa I}/n^{I}_{max}>\zeta$, the  natural states are defined by:
\begin{equation}
|\kappa^{IM}\rangle=\sum_{\beta\gamma\rho K}\frac{u^{\kappa IK}_{\rho}(\beta,\gamma)}{\sqrt{n^{\kappa I}}}| IMK,N,\beta,\gamma\rangle_{\rho}.
\label{natstates}
\end{equation}
Obviously, a cutoff  $\zeta$ has to be introduced in the value  of the norm eigenvalues to avoid linear dependences \cite{RingAMP_Rel_09}.
Then, the HWG equation is transformed into a normal eigenvalue problem:
\begin{equation}\label{HWG_nat}
\sum_{\kappa'}\langle\kappa^{I}|\hat{H}|\kappa'^{I}\rangle g^{\sigma I}_{\kappa'}=E^{\sigma I}g^{\sigma I}_{\kappa}.
\end{equation}
From the coefficients $g^{\sigma I}_{\kappa}$ we can define the so-called collective wave functions 
\begin{equation}
p^{\sigma I}_{\rho K}(\beta,\gamma)=\sum_{\kappa }g^{\sigma I}_{\kappa}u^{\kappa IK}_{\rho}(\beta,\gamma) 
\label{coll_wf}
\end{equation} 
that satisfy 
\begin{equation}
\sum_{\beta\gamma\rho K} |p^{\sigma I}_{\rho K}(\beta,\gamma)|^2=1,  \;\;\; \forall \sigma,
\label{norm_coll_wf}
\end{equation}
and are equivalent to a probability amplitude. 

  In our calculations we use the separable pairing plus quadrupole Hamiltonian with the same interaction strengths  as in  Ref.~\cite{Fang-Qi_2016}. These strengths have been fitted to reproduce the experimental deformation of the Erbium isotopes \cite{beta_exp} and to reproduce reasonable one dimensional potential energy surfaces.
  We consider  three major shells namely  $N=4,5,6$ for neutrons and $N=3,4,5$ for protons, the single particle energies are the same the ones used in Ref.~\cite{Fang-Qi_2016}.

\section{Results and discussion}\label{sect3}

We apply our theory to the calculation of collective bands in the Erbium isotopes $^{156-172}$Er, ranging from the soft ones (with neutron number $N=88$ or $90$) to the well deformed ones (up to $N=104$).  We solve the constrained HFB equations, Eqs.~(\ref{HFB_Eq}-\ref{Lagr}), in a triangular mesh of 49 points in the interval $0\leq \beta\leq 0.6$, $0^\circ \leq \gamma\leq 60^\circ$.


\subsection{Potential energy surfaces}
The intrinsic HFB states $| \Phi_{0}(\beta,\gamma)\rangle$, solution of Eq.~(\ref{HFB_Eq}), are not eigenstates of the 
symmetry operators. A PN and AM symmetry conserving state is provided by 
\begin{equation}\label{eq:proj_int}
 |IMK,N,\beta,\gamma\rangle_{0}= \hat{P}^{I}_{MK}\hat{P}^{N}\hat{P}^{Z} | \Phi_{0}(\beta,\gamma)\rangle,
 \end{equation}
 introduced in Eq.~(\ref{eqwf}).  They are eigenstates of the operators $\hat{I}^2,\hat{I}_{z},\hat{I}_{3},\hat{Z}$ and $\hat{N}$ and are the building blocks of our theory. The associated laboratory system state is provided by
 \begin{equation}\label{eq:proj_lab_0}
|\sigma,IM,\beta,\gamma \rangle_{0} =  \sum_{K} h^{\sigma IK}(\beta,\gamma)|IMK,N,\beta,\gamma\rangle_{0},
\end{equation}
and the coefficients $h^{\sigma}_{K}(\beta,\gamma)$ are determined by the variational principle. Minimisation of the energy with respect to the coefficients $h^{\sigma}_{K}$ provides a reduced HWG like equation
\begin{equation}
\sum_{K'}\left(\mathcal{H}^{IKK'}_{00}(\beta\gamma, \beta\gamma)-E^{\sigma I}(\beta,\gamma)\mathcal{N}^{IKK'}_{00}(\beta\gamma, \beta\gamma)\right)h^{\sigma IK'}(\beta\gamma)=0.
\label{HWG_eq_red}
\end{equation}
This simplified HWG equation is solved in the same way as the the full HWG equation. The index  $\sigma=1,2, ...$ numbers the $(2I + 1)$ different states that can be obtained with the angular momentum projection. However, because of the time reversal and the 
$e^{i\pi \hat{J}_z}$ symmetries  imposed on the intrinsic wave functions \cite{PSMreview}, this number is reduced to $(I/2+1)$ and $(I-1)/2$ states for even and odd values of $I$, respectively. Moreover, if we furthermore have axial symmetry, only one state can be obtained and only for even values of $I$. The eigenvalues $E^{\sigma I}(\beta,\gamma)$ provide the potential energy surfaces as a function of $(\beta,\gamma)$. The weights $h^{\sigma IK}(\beta\gamma)$ of Eq.~\ref{eq:proj_lab_0} allow to know the $K$ composition of the given state $I_{\sigma}$.  
For $I=0$ there is only one state and we display $E^{\sigma=1, I=0}(\beta,\gamma)$
 in Fig.~\ref{Erpes0}  as a function of $(\beta,\gamma)$
 for the  $^{156-172}$Er nuclei. 
As an overview of the results we obtain around the energy minimum flat triaxial  PES  for the lighter isotopes and  very steep, almost axially symmetric ones for the heavier ones.
In the first row, left panel, we display the nucleus $^{156}$Er, whose energy minimum is located at  $(\beta= 0.179, \gamma= 21.5^\circ)$, see Table~\ref{table1}. This triaxial nucleus is rather soft in the area $0.1\leq \beta \leq 0.25$, $0^\circ\leq \gamma \leq 60^\circ$. For larger $\beta$-values the energy increases steeply.  The nucleus $^{158}$Er, with deformation parameters $(\beta= 0.245, \gamma= 15.5^\circ)$, has a larger $\beta-$ and a smaller $\gamma$ value than the lighter $^{156}$Er. As a matter of fact the PES is very similar to the one of $^{156}$Er just shifted to larger $\beta$ deformation.
In the case of $^{160}$Er, $(\beta= 0.279, \gamma= 13.6^\circ)$, the shift to larger $\beta$ deformation 
and smaller $\gamma$ value continues but now the PES starts to be somewhat steeper and the area within the 1 MeV contour starts to shrink.  For the nuclei  $^{162-164}$Er the pace of shifting to larger  $\beta$ deformation slows down but the PESs 
becomes steeper with increasing  $\gamma$ values. In the second row the nuclei  $^{166-172}$Er are
displayed. Here the tendency observed in  $^{162-164}$Er continues but now in a  more moderate way.
\begin{figure*}[htbp]
	\includegraphics[width=1.0\textwidth]{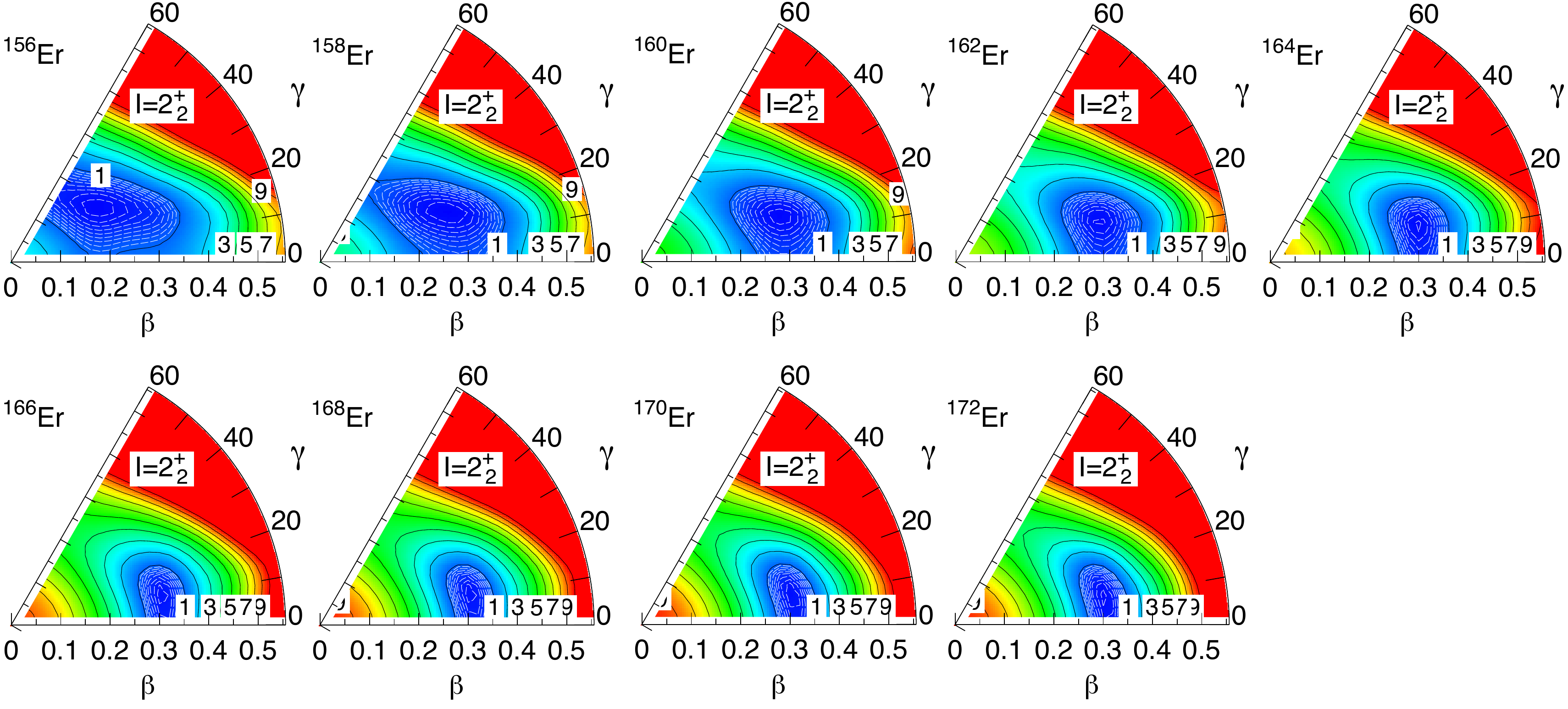}
	\caption{ (Color online) Particle number and angular momentum projected potential energy surfaces  for $^{156-172}$Er for $I=2^{+}_{2}$ in the $(\beta,\gamma)$ plane. The units for the contours are MeV and for $\gamma$ degrees. The energy origin has been set to zero at the energy minimum in each panel. The black continuous contour lines are 1 MeV apart and the white dashed lines around the minimum are separated by 0.1 MeV.}
	\label{Erpes2}
\end{figure*}

We turn now to the case $I=2$. In this case Eq.~(\ref{HWG_eq_red}) has two solutions $2_1$ and $2_2$. Both of them can have $K=0$ and $K=\pm2$ components in their wave functions.  In general, however, the lowest one, $2_1$, is a rather pure $K=0$ and $2_2$ a rather pure $K=2$.  The state $2_1$ is identified as the $I=2 \hbar$ member of the band built on the deformed intrinsic state $|\Phi_{0}(\beta,\gamma)\rangle$.  The state $2_2$ on the other hand is interpreted as the band head of the $\gamma$-band built on the same intrinsic state.   The PESs of the states $2_1$  are very similar to the ones of $0_1$ of Fig.~\ref{Erpes0} and will not be shown.  The PESs of the $2_2$ states for the $^{156-172}$Er nuclide are displayed in Fig.~\ref{Erpes2}.   A look at this figure and at Table~\ref{table1} reveals that the $\beta$ and $\gamma$ values of the energy minimum of the  $\gamma$ band head are larger, specially for the lighter isotopes,  than the ones of the ground state.    The largest differences between the PESs of $0_1$ and $2_2$ concern the $\gamma$ degree of freedom. We find that, even for the heaviest isotopes where the PESs are more rigid,  the $\gamma$ values of the $2_2$ energy minima are clearly larger than for the $0_1$. Another remarkable
 feature is that the $\gamma$-PESs are much softer in the $\gamma$ direction than the $0_1$.  

\begin{table}[htbp]
\centering
\caption{$\beta, \gamma$ values of the energy minimum of the potential energy surface for $I=0_1$ and $I=2_{2}$.}
\begin{tabular}{|c|cc|cc|}
  \hline
    & $\;\;\;\;\;\; I\;=$ &$ \! \!\!\!\!\!\!\!\!0_1$ & $\;\;\;\;\;\; I\;=$ & $\! \!\!\!\!\!\!\!\!2_2$ \\
    \hline
 &      $\beta$ & $\gamma$ & $\beta$ & $\gamma$\\
  \hline
   $^{156}$Er & 0.179 & 21.53 & 0.200 & 31.29 \\
  $^{158}$Er & 0.245 & 15.52 & 0.288 & 19.48 \\
  $^{160}$Er & 0.279 & 13.60 & 0.298& 17.27 \\
  $^{162}$Er & 0.290 & 11.55 & 0.306 & 16.79 \\
  $^{164}$Er & 0.299 & 11.22 & 0.308 & 15.22 \\
  $^{166}$Er & 0.302 &  9.60 & 0.313 & 12.09 \\
  $^{168}$Er & 0.305 &  8.07 & 0.313 & 12.09 \\
  $^{170}$Er & 0.310 &  9.33 & 0.311 & 13.61 \\
  $^{172}$Er & 0.305 &  8.07 & 0.305 & 12.43 \\
  \hline
\end{tabular}
\label{table1}
\end{table}


\subsection{Spectra}
The next step is the solution of the Hill-Wheeler equation, Eq.~(\ref{HWG_nat}), corresponding to the ansatz of Eq.~(\ref{eqwf}) which provides the energies and wave functions of the ground and excited states. Concerning the number of two-quasiparticle states considered in the GCM ansatz of Eq.~(\ref{eqwf}),  in this calculation we set the following energy cutoff condition:
\begin{equation}
E_{i}(\beta,\gamma)+E_{j}(\beta,\gamma)+E_{0}(\beta,\gamma)\leq E(\beta_{min},\gamma_{min})+3.0~{\rm MeV}
\end{equation}
where $E_{i}(\beta,\gamma)$ and $E_{j}(\beta,\gamma)$ represent the quasiparticle energies of the states  $i$ and $j$.  $E_{0}(\beta,\gamma)$ is the energy of the HFB state with deformation $(\beta,\gamma)$ given by
\begin{equation}
E_0(\beta,\gamma) = \langle \Phi_{0}(\beta,\gamma)| \hat{H} | \Phi_{0}(\beta,\gamma)\rangle,
\label{E_ground}
\end{equation}
with $| \Phi_{0}(\beta,\gamma)\rangle$ the solution of Eq.~(\ref{HFB_Eq}) and $(\beta_{min},\gamma_{min})$ represent the deformation of the HFB minimum.
$E(\beta_{min},\gamma_{min})$ is the energy minimum of the potential energy in the $(\beta,\gamma)$ plane. 
The convergence with this cutoff has been checked and  is  good. 

\begin{figure}[htbp]
\includegraphics[width=0.47\textwidth]{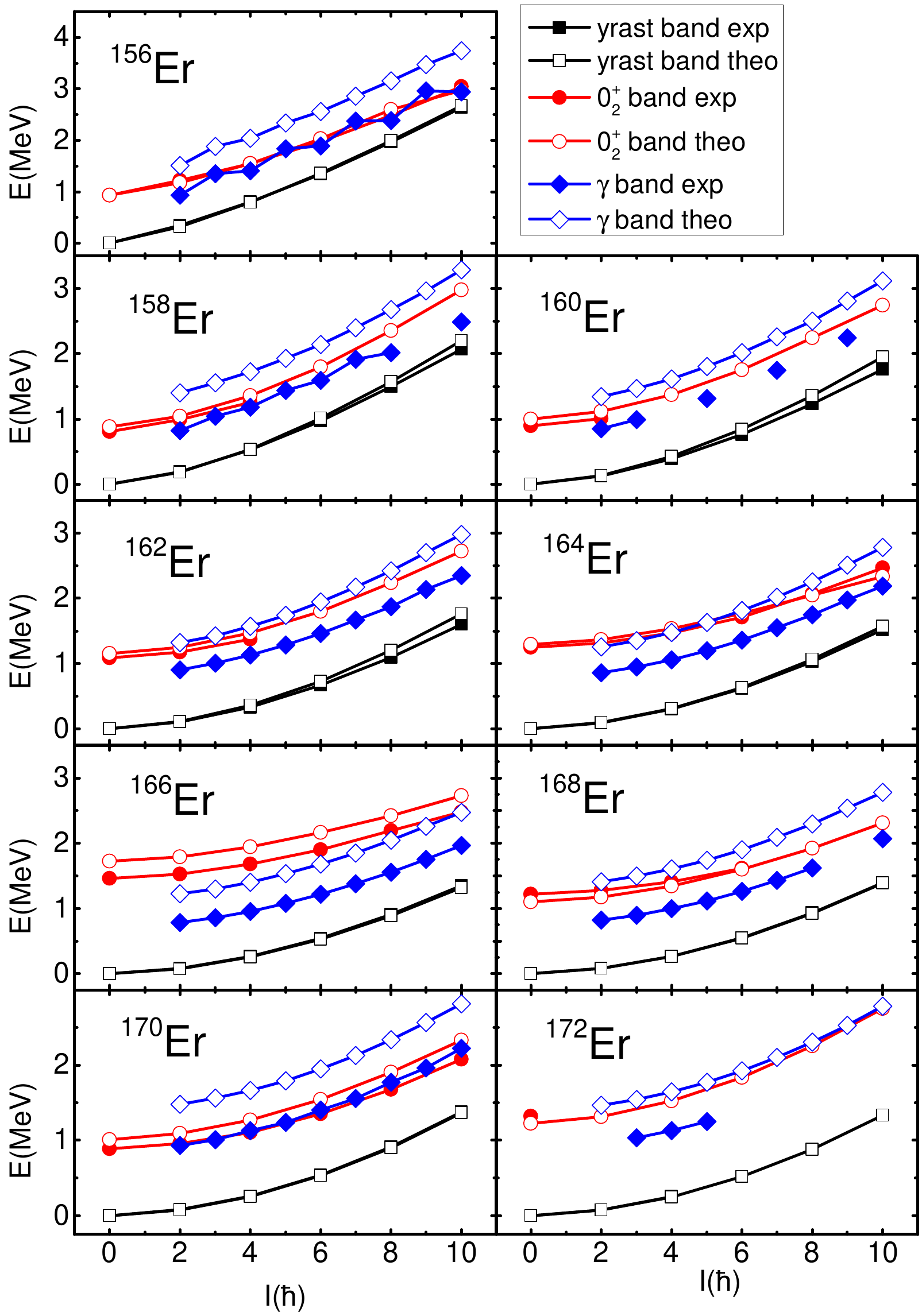}
\caption{(Color online) The calculated spectra of the yrast bands, the $0^{+}_{2}$ bands and the $\gamma-$bands of the $^{156-172}$Er nuclei, compared with the experimental data. The data were taken from \cite{ENSDF}. For most of the isotopes the theoretical values for the Yrast band are on top of the experimental ones.}
\label{Erspectrum}
\end{figure}

In Fig.~\ref{Erspectrum}  we display the energies $E^{\sigma I}$,  the eigenvalues of Eq.~(\ref{HWG_nat}),  for the collective bands of the  $^{156-172}$Er isotopes (alternatively we will also denote these states by $E^{I_\sigma}$), namely  the Yrast band, the band based on the $0^{+}_{2}$ state, also called $\beta-$band, and the $\gamma-$band. For the identification of the bands we have used the transition probability as well as the $K$ distribution of the corresponding wave functions.   Notice that the high spin members of the $0^{+}_{2}$ bands and the $\gamma-$bands do not necessarily coincide with the lowest excited state with angular momentum $I$, i.e. with $I^{+}_{2}$ state.  For the excited bands and for very high angular momentum, the band structure may be dominated by two-quasiparticle and four-quasiparticle states. At present the four-quasiparticle configurations are not included in our model space, therefore we only show these bands up to $I=10\hbar$. Concerning the Yrast bands our theoretical results (black empty squares) are in excellent agreement with the experimental data (black filled squares).  For most  isotopes the theoretical values are on top of the experimental ones. It is interesting to notice that these results are close to the ones obtained in the axially symmetric calculations of Ref.~\cite{Fang-Qi_2016} in spite of the fact that all of them have a triaxial minimum in the PESs, see Fig.~\ref{Erpes0}. We will return to this point later on in the discussion of the wave functions.  The small changes in the wave function obtained by the consideration of the $\gamma$ degree of freedom improve the results of Fig.~\ref{Erpes0}. 
The experimental (red filled circles) and the theoretical values (red empty circles) of the $0^{+}_{2}$ bands are also plotted in Fig.~\ref{Erspectrum}. The  agreement is again very good for most nuclei. In comparison with the axially symmetric calculations of Ref.~\cite{Fang-Qi_2016} in the $\gamma$ soft nucleus $^{156}$Er  the agreement  between theory and experiment
has been considerably improved by the inclusion of the $\gamma$ fluctuations in the calculations. It should also be noticed that the moments of inertia of the $0^{+}_{2}$ bands are also well reproduced, especially in $^{164-168}$Er, in which the $0^{+}_{2}$ bands have larger moments of inertia than the yrast bands.
We now turn to the $\gamma$ band.  The experimental values are represented by blue filled rhombi and the theoretical values by blue empty rhombi. For this band the agreement between theory and experiment 
is not as good as for the former bands. The theoretical values are higher than the experimental ones.
But since the experimental moments of inertia are very well reproduced  by the theory for all isotopes the real problem are the band head energies which are too high. 
  This is a common feature of many theoretical approaches, like the Quasiparticle Random Phase Approximation with the Skyrme force \cite{Tera} or the 5DMCH \cite{Bohr_5DM} with the Gogny force. In particular, in the latter publication after a study of 354 nuclei the authors conclude that the theoretical energies of the band heads of the $\gamma-$bands are on average about $25\%$  higher than the experimental ones. Another  aspect is the staggering of the $\gamma-$band. As in the experimental values we obtain staggering in the soft $^{156}$Er and $^{158}$Er nuclei and no staggering in the heavier rigid ones. The phase of the staggering in the theory is also the same as in the experiment but the theoretical amplitudes are much smaller than the experimental ones.

From the discussion of the energies of the collective bands we conclude that the yrast states and the excitation energies of the $0^{+}_{2}$ bands are  well reproduced as shown in Fig.~\ref{Erspectrum}.  Based on the good agreement with the experimental results it seems that in these Er isotopes, the important degrees of freedom are the shape vibration and the two-quasiparticle states. 
  The reason why the band heads of the gamma band are higher than in the experimental data is not clear to us.  We  think that our theory contains enough degrees of freedom to describe correctly the gamma band, and that, though more realistic interactions  also provide too high energies for the gamma band heads, in our case the reason must be looked for either in the single particle energies or in the simplicity of the pairing plus quadrupole interaction.  
  
\begin{figure}[htbp]
\centering
\includegraphics[width=0.45\textwidth]{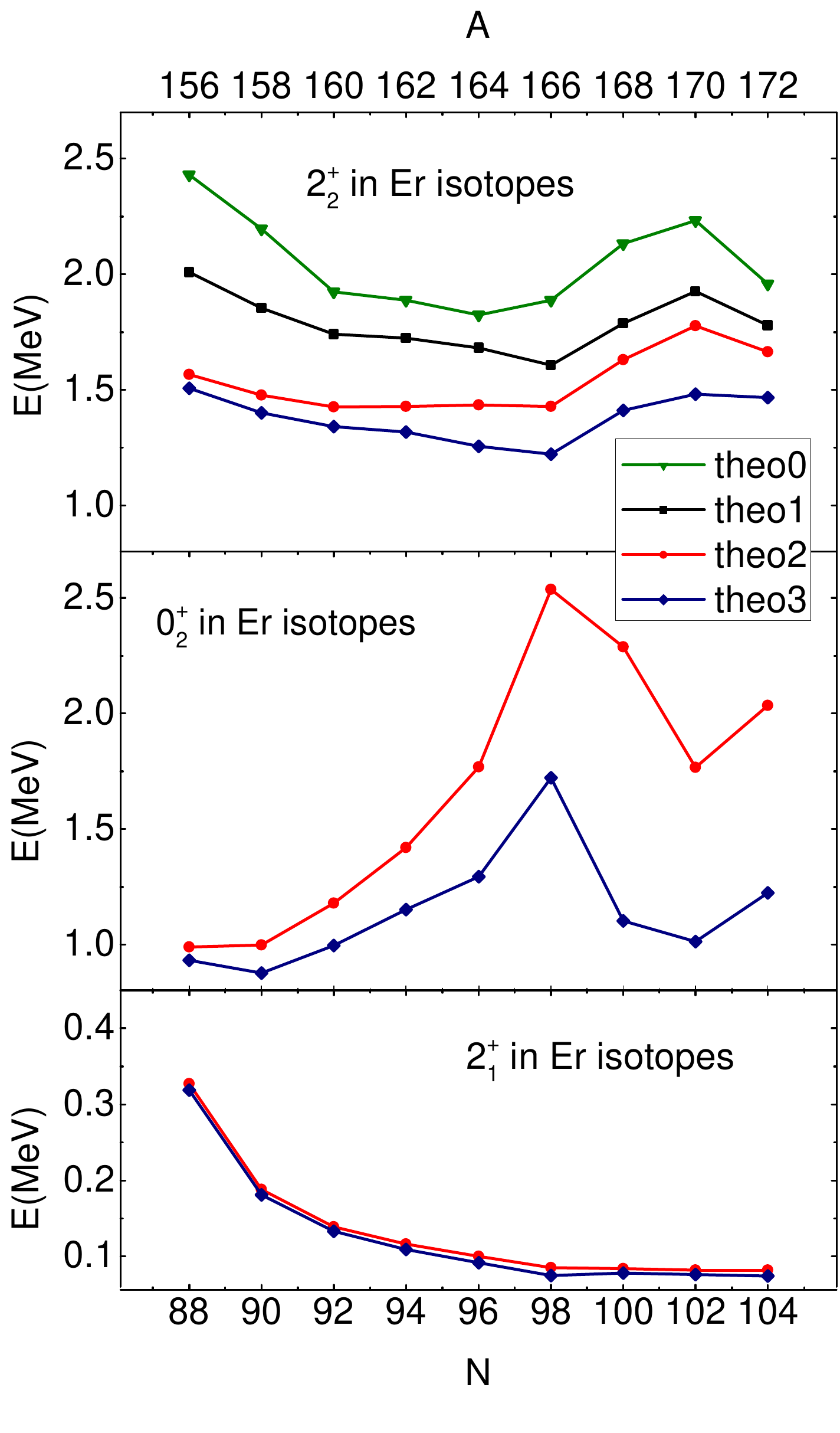}
\caption{ (Color online) Top panel: Energies of the band head of the $\gamma$-band for the Erbium isotopes calculated according to Eq.~(\ref{eq:theo0}) (``theo0"), Eq.~(\ref{eq:theo1}) (``theo1"), Eq.~(\ref{eqwf1}) (``theo2") and Eq.~(\ref{eqwf}) (``theo3"). Middle  panel: Energies of the band head of the $\beta$-band for the Erbium isotopes calculated according ``theo2" and ``theo3".   Bottom panel: Energies of the $2^{+}_{1}$ states according to ``theo2" and ``theo3".}
\label{GCMvs2QP}
\end{figure}

   In our approach, see Eq.~(\ref{eqwf}), we have two main degrees of freedom, namely the shape vibrations and the two-quasiparticle excitations. To disentangle the different contributions to the final energies displayed in Fig.~\ref{Erspectrum} we  have performed different calculations for the band heads of the collective bands.  
     Let us first discuss the  behaviour of the $\gamma$ band heads with the neutron number.  A simple description of the $\gamma$-band can be obtained ignoring shape fluctuations and quasiparticle degrees of freedom  within the projected mean field approach.  First, we consider the energy minimum of the PES displayed  in Fig.~\ref{Erpes0}, which we shall call $(\beta_{\rm min}^{1,0},\gamma_{\rm min}^{1,0})$, where the superscripts ${1,0}$ stand for ${\sigma=1, I=0}$.  Its energy value is denoted by $E^{0_{1}} (\beta_{\rm min}^{1,0},\gamma_{\rm min}^{1,0}) \equiv E^{\sigma=1,I=0}(\beta_{\rm min}^{1,0},\gamma_{\rm min}^{1,0})$, see Eq.~(\ref{HWG_eq_red}). Second, solve Eq.~(\ref{HWG_eq_red}) for $I=2$ at the $(\beta_{\rm min}^{1,0},\gamma_{\rm min}^{1,0})$ point.  The energy of the gamma band head is given by 
 \begin{equation}  
  E(2^{+}_{\gamma})= E^{\sigma=2,I=2}(\beta_{\rm min}^{1,0},\gamma_{\rm min}^{1,0})- E^{\sigma=1,I=0}(\beta_{\rm min}^{1,0},\gamma_{\rm min}^{1,0}).
 \label{eq:theo0}
\end{equation}    
This zero order approach has been denoted "theo0" in Fig.~\ref{GCMvs2QP} and is represented with green filled down-triangle  symbols in the top panel.  This approach assumes that the $\gamma$ band head  and the ground state have the same minimum.  It predicts too high values for the $\gamma$ band heads, there is a kind of plateau for $N=92, 94$ and 96 with rising energy values at both sides of the plateau.

The next degree of sophistication still considers a projected mean field wave function for the  $\gamma$ band but it includes  the possibility of a different mean field for the $\gamma$-band and the ground band. To find this state one  explores  the $(\beta,\gamma)$ plane PES of Fig.~\ref{Erpes2} to look for a lower $\gamma$ band head. Using a similar notation this approach provides 
 \begin{equation}  
E(2^{+}_{\gamma})=   E^{\sigma=2,I=2}(\beta_{\rm min}^{2,2},\gamma_{\rm min}^{2,2})- E^{\sigma=1,I=0}(\beta_{\rm min}^{1,0},\gamma_{\rm min}^{1,0}),
 \label{eq:theo1}
\end{equation}   
where  $(\beta_{\rm min}^{2,2},\gamma_{\rm min}^{2,2})$ denotes the location of the energy minimum of the PES of Fig.~\ref{Erpes2}.
This approach has been denoted "theo1" in Fig.~\ref{GCMvs2QP} and is represented by black  filled squares. As we can see in the top panel of this figure it produces a general lowering of the $\gamma$ energy for all isotopes of about 250-500 KeV. 
This approach, though considering different intrinsic shapes for the ground state and the $\gamma$ band head, does not mix shapes.
   Our third approach  includes HFB vacua of all deformations allowing for shape mixing but does not include any two-quasiparticle state, i.e. the ansatz of Eq.~(\ref{eqwf}) is replaced by
\begin{eqnarray}\label{eqwf1}
|\sigma,IM \rangle&=&\int d\beta d\gamma \sum_{K} f^{\sigma,IK}_{0}(\beta,\gamma) \hat{P}^{I}_{MK}\hat{P}^{N}\hat{P}^{Z} |\Phi_{0}(\beta,\gamma)\rangle.
\end{eqnarray}
This approach is commonly used to describe collective motion in the scientific literature, in particular $\gamma$-bands. It has been denoted "theo2" in Fig.~\ref{GCMvs2QP} and is represented by red circles. If we compare this approach with "theo0" we observe an almost constant lowering of 500 keV for most isotopes with the exception of the soft nuclei $^{156}$Er and $^{158}$Er where we obtain about 850 and 750 keV respectively.  These results clearly illustrate the relevance of shape mixing for the description of $\gamma$-bands.

Our last approach  corresponds to the full ansatz of Eq.~(\ref{eqwf}) and includes the HFB vacua of all deformations as well as two-quasiparticle states built on each of them and allows for all kind of mixing among all states.  This approach has been denoted "theo3" in Fig.~\ref{GCMvs2QP} and is represented by blue filled up-triangle  symbols in the top panel. In this case
we observe that, as compared with the previous case,  the mixing of quasiparticles produces smaller energy lowering. 
Interestingly for the softer light nuclei the changes are much smaller than for the rigid heavier ones.  Combining the results of the last two approaches one can conclude that, as expected, shape fluctuations are more relevant for soft nuclei and quasiparticle degrees of freedom for rigid ones. 

\begin{figure*}[htbp]
\centering
\includegraphics[width=1.0\textwidth]{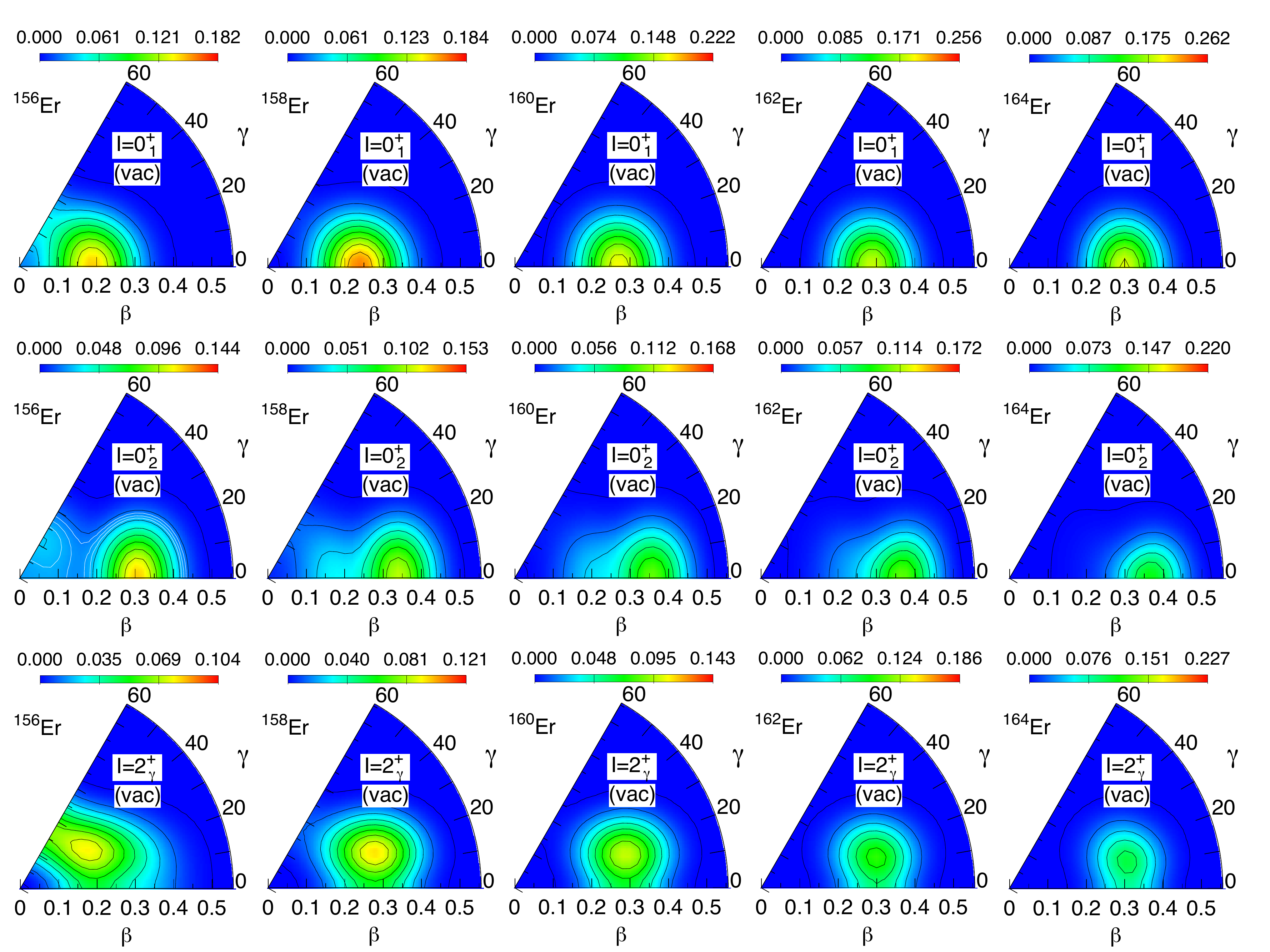}
\caption{ (Color online) Vacua probability amplitude  $\mathcal{P}^{\sigma I}_{\rm \!\!vac}(\beta,\gamma)$, Eq.~(\ref{vac_coll_wf}), in the $(\beta,\gamma)$ plane, calculated for the $0^{+}_{1}$  states (top panels), the $0^{+}_{2}$  states (middle  panels) and the $2^{+}_{\gamma}$  states (bottom  panels)  in $^{156-164}$Er. 
	The units  for $\gamma$ are degrees.  
	The maximum of the wave function is indicated in the palette on top of each panel. The contour lines start at zero and increase by a tenth of the corresponding maximal value from one contour to the next. (See the text)}
\label{wf_156-164_vac}
\end{figure*}

We now turn to the discussion of the different contributions for the $\beta$-band, see the middle panel of Fig.~\ref{GCMvs2QP}. Unfortunately the first two approaches 
discussed for the $\gamma$-band  are not possible for the $0_{2}$ band because for $I=0$  the HWG equation, Eq.~(\ref{HWG_eq_red}), provides only one solution at each point of the $(\beta,\gamma)$ plane and since there is only one minimum in the PES of  Fig.~\ref{Erpes0}, this corresponds to the $0_1$ state.  The results for the last two approximations, namely "theo2" and "theo3" are presented in the bottom panel of Fig.~\ref{GCMvs2QP}.  
As for the $\gamma$-band the mixing of quasiparticle states in the GCM ansatz modifies very little the energies of the 
$\beta$ band head for the light isotopes (soft nuclei) and it produces a larger energy lowering for the heavier rigid ones.
Though in the $\gamma$-band case one observes a lowering of the energy with increasing neutron number, it is much more pronounced in the case of the $\beta$-band. That means the quasiparticle degrees of freedom are much more relevant for the $\beta$-band than for the $\gamma$-band. 

The general behaviour of the band heads of the $\beta$ and  $\gamma$ bands as a function of the neutron number is  different for both bands. The $\beta$ band head energies present a maximum at $N= 98$, a marked decrease up to $N=102$ and an increase towards $N=104$ whereas  the  $\gamma$ band energies  display a minimum at $N= 98$  and then an smooth increase up to  $N=104$.  
The structure of the $\beta$ band heads can be understood in general terms: looking at a Nilsson diagram we find a deformed shell closure at $N=98$ and a less pronounced one at $N=104$. Since for shell closures one expects higher quasiparticle energies and since these are very relevant for the description of the $\beta$-band, the mentioned closures roughly explain the observed behaviour.

Concerning the structure of the $\gamma$ band heads as a function of $N$  we do expect 
a smoother dependence with $N$ than for the $\beta-$band since in the former the quasiparticles play a much smaller role than in the latter.

 Lastly, the  behavior of the $2^{+}_{1}$ states is displayed in the bottom panel of Fig.~\ref{GCMvs2QP}. Here we can observe that the two-quasiparticle states  produce a small decrease   of the excitation energy of the  $2^{+}_{1}$ states.  This decrease, however, is much smaller than the ones observed for the heads of the $\beta$ and $\gamma$ bands.

 In BMFT calculations of light nuclei with effective forces like Skyrme, Gogny or relativistic ones, and with only shape fluctuations, a stretching of the spectrum has been observed, see for example Ref.~\cite{RE.07}. Several reasons have been argued for this behavior, among others the absence of 
	two-quasiparticle states in the calculations, the lack of an angular momentum projection before the variation or the need to break the time reversal 
symmetry. The first two issues are difficult to implement with effective forces because of the amount of CPU time needed.  The third one, however, has been  carried through in Refs.~\cite{BRE.15,EBR.16} and a significant compression of the spectra has been obtained.   As we have seen in the preceding discussion the consideration of 2qp  states also produces a compression of the spectra.   It will be interesting to check if the breaking of the time reversal symmetry in the present calculations will bring the results of the $\gamma$ band to a better agreement with the experimental results. 


\begin{figure*}[htbp]
\centering
\includegraphics[width=0.99\textwidth]{{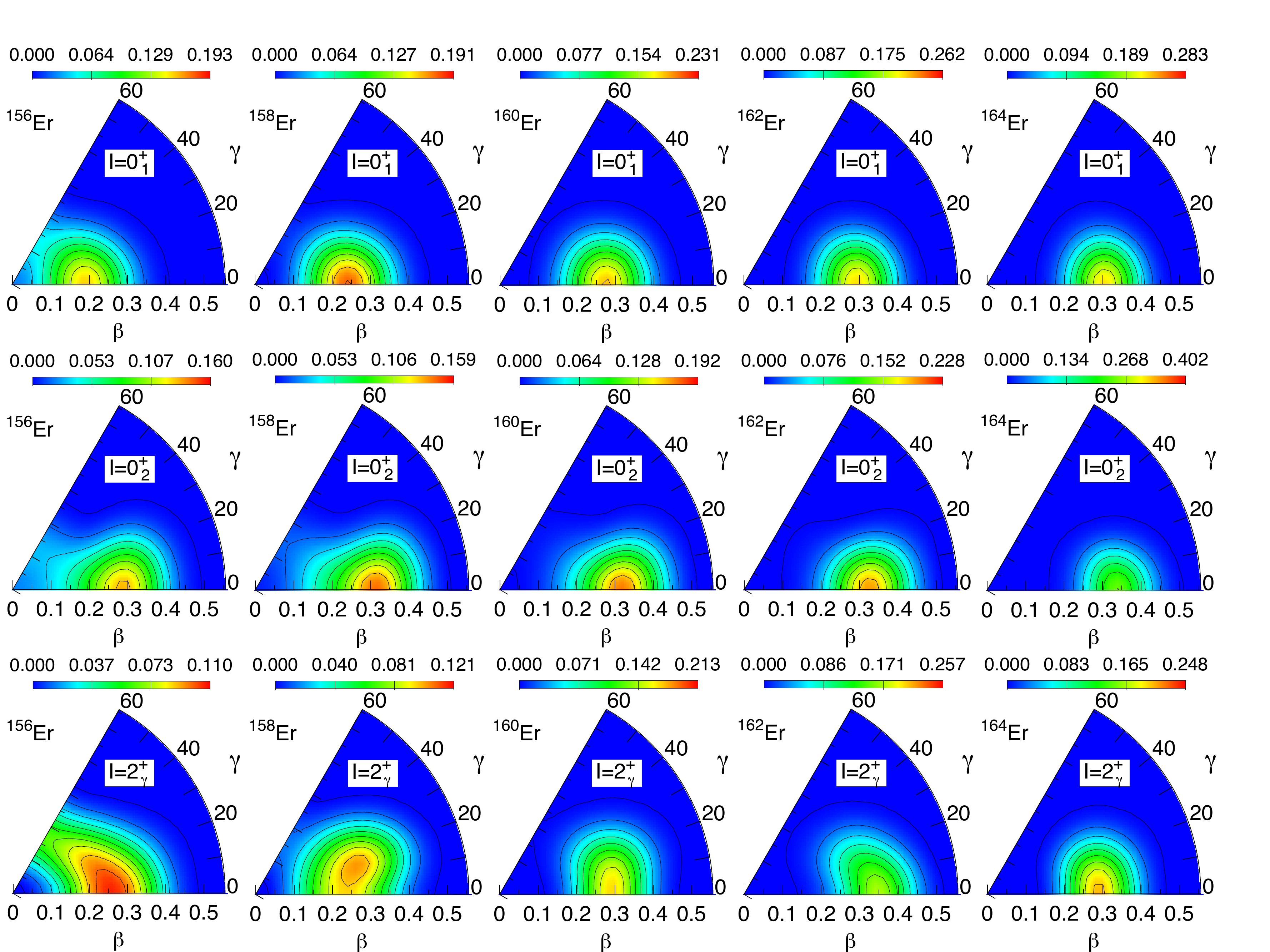}}
\caption{ (Color online)  Same as Fig.~\ref{wf_156-164_vac} but for the total  probability 
amplitude distribution $\mathcal{P}^{\sigma I}(\beta,\gamma)$, Eq.~(\ref{tot_coll_wf}).}
\label{wf_156-164_all}
\end{figure*}

\subsection{Collective wave functions}

For a better understanding of the results it is convenient to analyse the wave functions
of the different states and approximations.
  
 Based on the collective wave function,  see Eq.~(\ref{coll_wf}), we can define additional quantities which 
provide relevant  information about the $(\beta,\gamma)$ distribution, thus
\begin{equation}
\mathcal{P}^{\sigma I}(\beta,\gamma)= \sum_{K \rho} |p^{\sigma I}_{\rho K}(\beta,\gamma)|^2 
\label{tot_coll_wf}
\end{equation}
gives the total weight of the $(\beta,\gamma)$ point in the wave function.  In the same way
\begin{equation}
\mathcal{P}^{\sigma I}_{\rm \!\!vac}(\beta,\gamma)= \sum_{K} |p^{\sigma I}_{0 K}(\beta,\gamma)|^2 
\label{vac_coll_wf}
\end{equation}
and
\begin{equation}
\mathcal{P}^{\sigma I}_{\rm \!\!2qp}(\beta,\gamma)= \sum_{K(i,j)} |p^{\sigma I}_{(ij) K}(\beta,\gamma)|^2 
\label{2qp_coll_wf}
\end{equation}
provide the weights of the vacua $|\Phi_{0}(\beta,\gamma)\rangle$ and the two-quasiparticle states $|\Phi_{ij}(\beta,\gamma)\rangle$, respectively, in the total wave function of the state $I_{\sigma}$.   This separation makes sense because it allows to differentiate  the contribution of the collective degrees from the single particle ones.

We concentrate on the wave functions of the band heads of the Yrast, the $\beta$ and  $\gamma$ bands. For pedagogical reasons we first discuss the probability amplitude of the vacua, see Eq.~(\ref{vac_coll_wf}), in these bands.
 In Fig.~\ref{wf_156-164_vac} we plot  $\mathcal{P}^{\sigma I}_{\rm \!\!vac}(\beta,\gamma)$   in the $(\beta,\gamma)$ plane for the isotopes $^{156-164}$Er. The results for the nuclei $^{166-172}$Er have been omitted  because they are in the rigid deformation limit and they just
 peak at the minima of the PESs of Fig.~\ref{Erpes0} and look  similar to the ones of the  $^{164}$Er nucleus. 
In all panels the maximum of the plotted wave function is indicated in the palette on top of the corresponding panel. The contour lines start at zero and increase by a tenth of the corresponding maximal value from one contour to the next. In principle each panel should present 10 contours but this is not always the case because the grid is not very dense. It  is also due to the plotting program which in some cases, very sharp wave functions for example, is not able to determine the contours corresponding to the largest values. That means that the missing contours, if any,  are always the largest ones. In the top panels the probability amplitude associated to the wave functions of the ground state of the corresponding nuclei are presented. In the nucleus $^{156}$Er the maximum of  the probability amplitude appears close to $(\beta = 0.2, \gamma = 0^{\circ})$. If we look at Fig.~\ref{Erpes0} or Table~\ref{table1} we find that the potential energy minimum appears at $(\beta = 0.18, \gamma = 21.5^{\circ})$. That means that the shape and quasiparticle mixing drives the maximum of the probability amplitude distribution to prolate shapes. This is not very surprising since the PES of this nucleus is very flat towards prolate shapes. We also find that the probability amplitude distribution is soft towards oblate shapes. In the nucleus $^{158}$Er the maximum of the distribution appears close to $(\beta = 0.25, \gamma = 0^{\circ})$ at variance with the potential minimum values of $(\beta = 0.24, \gamma = 15.5^{\circ})$. We also observe  a decrease of the 
contour values on the oblate side.  
With increasing mass number the probability amplitude  of the ground states shifts to larger deformations loosing thereby softness in the $\gamma=60^{\circ}$ direction. We also observe that with growing neutron number the probability amplitude distributions get sharper. This can be easily realised noticing that the maximum value of the distribution increases (see the palettes) and that the contour lines corresponding to zero experience a contraction.   This is the behaviour one expects from a transition of a soft nucleus to a rigid one.   

In the second row we display the probability amplitude distribution of the vacua for the $\beta$ band heads of the Erbium isotopes.  As for the ground state  the distribution peaks on the prolate axis for all isotopes.
At first sight one observes clear differences between the probability amplitude distribution of this band head and the ground state.  As compared with the ground bands, the $\beta$-band distributions are broader for the  $^{156-162}$Er isotopes and similar for the $^{164-172}$Er isotopes, the heaviest ones not being shown here. For the lighter isotopes the $\beta$-band distributions are shifted to higher deformations and are in general flatter (see the maximum value on the palette of each panel) than in the ground state.  Though in principle one could think 
that the maximum of the distribution for the band head of the $\beta$-band should coincide with the one of the ground state, we see that this is not the case. The dynamic induced by the shape mixing shifts the maximum to larger deformations. This is not the case for the heavier isotopes$^{166-172}$Er.  There the PESs are very steep, see Fig.~\ref{Erpes0}, and the maximum of the wave function coincides with the minimum of the corresponding PES.

One also observes a somewhat strange behaviour of the probability amplitude distribution  at small deformations especially in the lighter isotopes.  In a genuine collective model, i.e.,  without coupling to other modes (for example triaxial deformation), the probability amplitude distribution of a $\beta$ vibration will display a zero as a function of the deformation parameter $\beta$, see for example Ref.~\cite{Fang-Qi_2016}. In Fig.~\ref{wf_156-164_vac} and for $^{156}$Er we have added three more contours in white colour to illustrate the fact that the probability amplitude distribution presents a minimum along the $\beta$-axis at $\beta \approx  0.15$, a valley in the $(\beta,\gamma)$ plane.  This minimum is a reminiscence of the zero that one would obtain without coupling to other modes. For slightly heavier isotopes, this minimum still persists, for example $^{158}$Er, but for even heavier ones it is washed out and only  long tails at smaller deformation remain.

The  probability amplitude  of the vacua for the $\gamma$ band heads is presented in the third row of Fig.~\ref{wf_156-164_vac}.  These distributions are a reflection of the PESs displayed in Fig.~\ref{Erpes2} as they should in a first approximation. The most genuine is the one of $^{156}$Er. The fact that the distribution presents a maximum at $(\beta=0.2, \gamma=30^{\circ})$ indicates that we have to deal with a quasi-$\gamma$ band $(n_\gamma=0)$ \cite{RS.80}. The distribution is broad and soft in the $(\beta,\gamma)$ degrees of freedom. The next isotope $^{158}$Er  presents its probability maximum at $(\beta=0.288, \gamma=20^{\circ})$. The shift to larger $\beta$-values as compared to $^{156}$Er causes a  considerable decrease in the  probability amplitude distribution close to the oblate axis. The distribution still is broad and collective (in the sense that many $(\beta,\gamma)$ values do have a non-zero probability amplitude).  With increasing neutron number the tendency is maintained, the maximum of the distribution shifts to larger $\beta$-deformation, the distribution becomes sharper and the collectivity diminishes. All this is very much in agreement with what one could expect from the PESs depicted
in Fig.~\ref{Erpes2}. It is interesting to realise that, in contrast with the results for the $\beta$-band, the maximum of the wave function is not much affected by the mixing with other shapes.

We now discuss briefly the total collective wave function $\mathcal{P}^{\sigma I}(\beta,\gamma)$, see Eq.~(\ref{tot_coll_wf}).  In Fig.~\ref{wf_156-164_all} we display these quantities in the same arrangement as in Fig.~\ref{wf_156-164_vac}.   The ground state probability amplitude distributions
depicted in the first row look very similar to the vacua distributions plotted in Fig.~\ref{wf_156-164_vac}.  The reason is rather simple, the two-quasiparticle contributions are very small as we can realise by comparing the maximum values of the corresponding probability amplitude in both plots.  That means two-quasiparticle states are not relevant for the description of the low spin members of the ground band.  In the second row we present now the results for the band heads of the $\beta$ bands.  We now find significant differences by comparison to the corresponding panels in Fig.~\ref{wf_156-164_vac}:  In general, the total distributions are shifted to smaller $\beta$ deformations, they are sharper and thereby a bit more concentrated. From these features we conclude that the contribution of the two-quasiparticle states is concentrated on the prolate side at smaller deformations than the ones favoured by the shape vibrations.  The largest change takes place for the softer isotopes $^{156-162}$Er, where the characteristic valley of the $\beta$-vibration is washed out considerably. That means, that the quasiparticle degrees of freedom, as well as the triaxial one,  couple
significantly with the genuine shape $\beta$-vibration.

The probability amplitude distribution of the total collective wave function of the $\gamma$ band heads is presented in the third row of Fig.~\ref{wf_156-164_all}.  We find that the maxima of the distribution do not correspond to the minima of the PESs of Fig.~\ref{Erpes2}, they are shifted to the prolate axis. They also differ from the corresponding probability amplitude distributions of the vacua 
shown in Fig.~\ref{wf_156-164_vac}. As in the $\beta$-band case the largest differences are observed for the softer isotopes  $^{156-160}$Er. The nucleus
$^{156}$Er presents a very soft distribution in the $\gamma$ direction, indicating a dominance of triaxial shapes, although it has its maximum close to the prolate axis. Its average deformation, as in the case of the ground state, is moderate. In $^{158}$Er  we find the maximum of the distribution at 
about $15^{\circ}$ but the softness in the $\gamma$ direction has decreased considerably. The heavier isotopes present as probability amplitude distributions stretched semicircles centred on the prolate
axis.  The differences between the vacua and the total distributions is due to the coupling of the quasiparticle degrees of freedom to the $\gamma$-vibration.

The two quasiparticle distributions $\mathcal{P}^{\sigma I}_{\rm \!\!2qp}(\beta,\gamma)$, Eq.~(\ref{2qp_coll_wf}), are not shown here. In general their probability amplitudes look like semicircles centred on the prolate axis at the point where the corresponding PES of 
Fig.~\ref{Erpes0} has  its maximum.  Looking at the plots of Fig.~\ref{wf_156-164_vac}-\ref{wf_156-164_all} it is easy to localize the centre of the distribution.


\subsubsection{Total 2-qp contribution to the wave function}
 \label{2qp_distri}

\begin{figure}[htbp]
\centering
\includegraphics[width=0.5\textwidth]{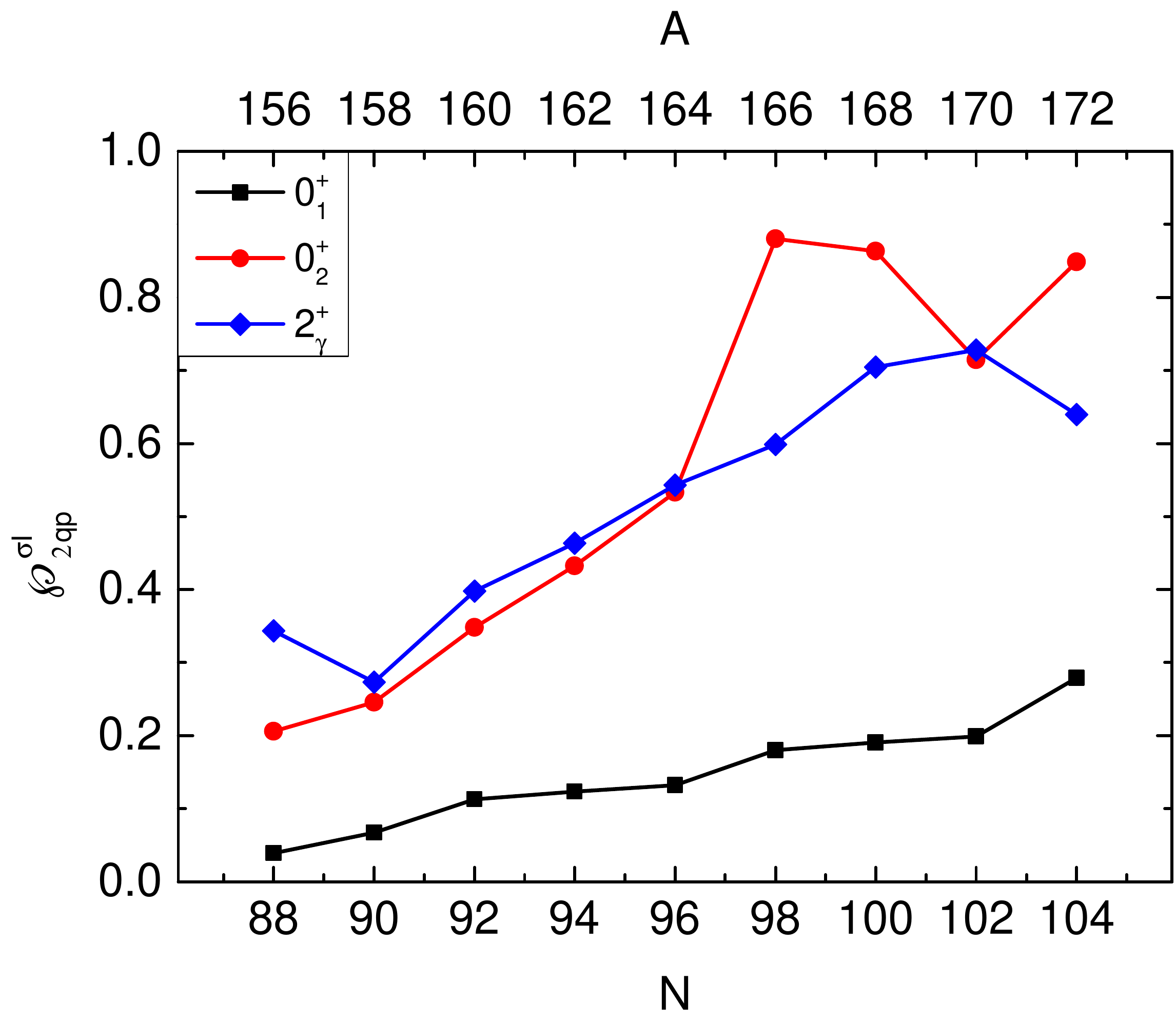}
\caption{ (Color online) Contribution of the two-quasiparticle states to the ground states and the $\beta$ and $\gamma$ band heads in $^{156-172}$Er, measured by the summation of the absolute square of the collective amplitudes over all two-quasiparticle states.}
\label{Er2qpweight}
\end{figure}

In Fig.~\ref{wf_156-164_vac} and Fig.~\ref{wf_156-164_all} we have presented the vacua and the total probability amplitude of the Erbium isotopes.  As mentioned  above the two-quasiparticle probability amplitude distributions as a function of the deformation parameters  $(\beta,\gamma)$ are rather similar.  However, we can quantify  the relevance of the two-quasiparticle excitations in the different band heads by calculating the total contribution of all two-quasiparticle configurations.
This term is given by 
\begin{equation}
 \mathcal{P}^{\sigma I}_{\rm \!\!2qp} = \sum_{\beta\gamma} \mathcal{P}^{\sigma I}_{\rm \!\!2qp}(\beta,\gamma),  
\label{2qp_tot}
\end{equation}
and plotted in Fig.~\ref{Er2qpweight} as a function of the neutron number.  
We find that the quasiparticle contribution to the wave function is different for the ground state and for the excited states. For the ground state we obtain always a modest contribution which
steadily increases with the neutron number.   For the $0^{+}_2$ states it starts already with a $20\%$ contribution and increases smoothly up to $N=96$ where it surpasses the $50\%$. 
The heavier isotopes $^{166-172}$Er reach, on the average, the  $85\%$ of the total wave function. This is what one would expect in the well deformed limit.
Concerning the quasi-$\gamma$ band head, it behaves rather similarly to the $\beta$-band up
to $N=96$, afterwards the percentage of two-quasiparticles in the wave functions increases 
smoothly and it reaches its maximum of about $70\%$ of the total wave function.

\subsubsection{K-Distribution of the wave functions}

\begin{figure}[htbp]
\centering
\includegraphics[width=0.5\textwidth]{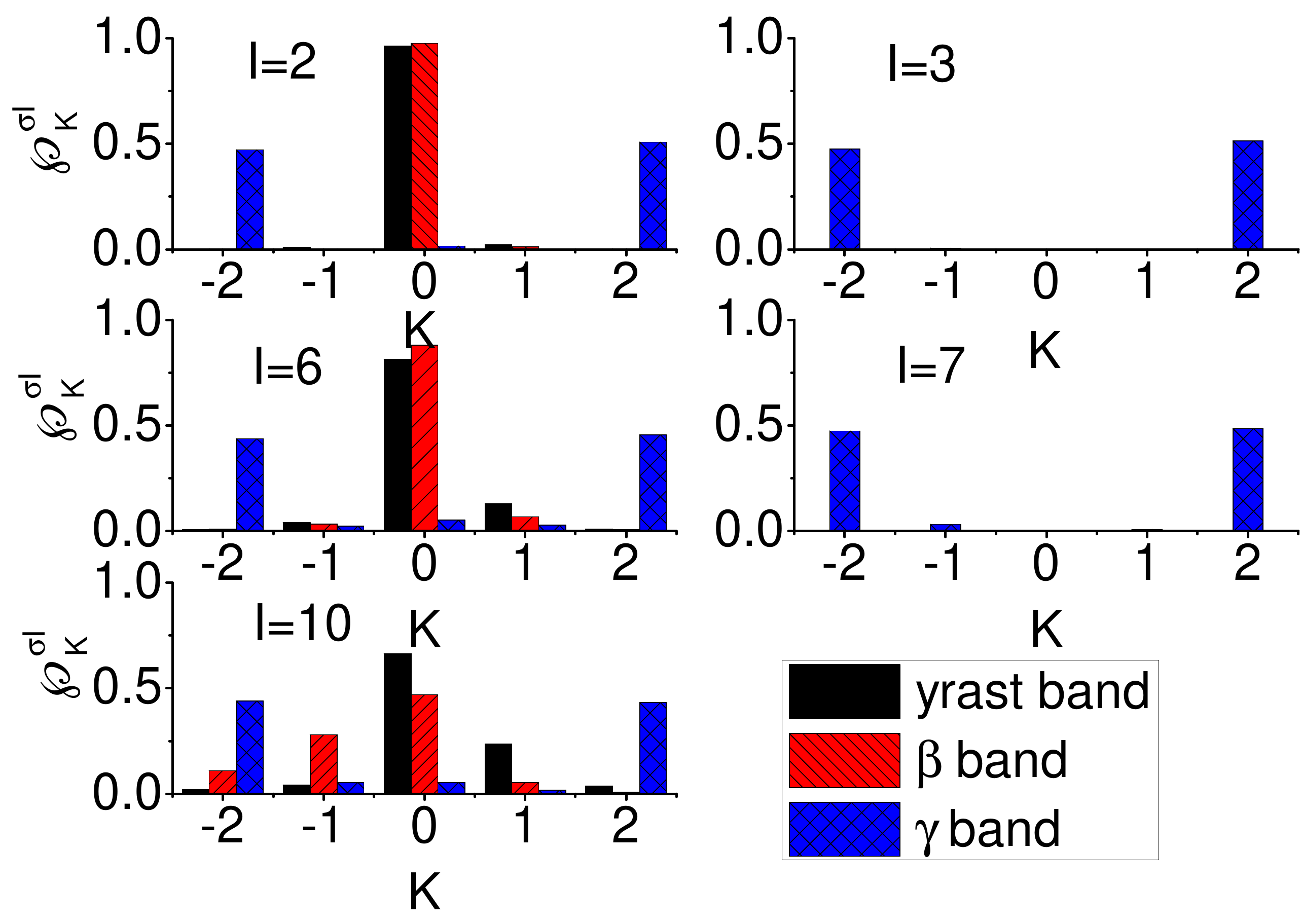}
\caption{ (Color online) K distribution  of the Yrast-, $\beta$- and $\gamma$-band of  $^{158}$Er for selected values of the angular momentum.}
\label{fig:K-distribution}
\end{figure}

A quantity interesting to characterise the wave functions is its K-distribution.
In terms of the quantities defined in the subsection above the  K-distribution is provided by
\begin{equation}
\mathcal{P}_{K}^{\sigma I} =\sum_{\beta\gamma\rho } |p^{\sigma I}_{\rho K}(\beta,\gamma)|^2. 
\label{K_coll_wf}
\end{equation}
 Since the plots look similar for different nuclei we only show the results for the nucleus   $^{158}$Er.   In Fig.~\ref{fig:K-distribution} we display $\mathcal{P}_{K}^{\sigma I}$ for several states of the Yrast-, $\beta-$ and $\gamma$-band which, together with the transition probabilities, have been used to identify the different bands. The left hand panels are  for the even spin values $I= 2 \hbar, 6 \hbar$ and $10 \hbar$ and the right ones for $I= 3 \hbar$ and $7 \hbar$.  In the histograms the black bars correspond to the Yrast band, the red to the $\beta$-band and the blue to the $\gamma$-band.  For each value of $K$ on the x-axis there are allotted three slots, the left one for the Yrast band, the middle one for the $\beta$-band and the right one for the $\gamma$-band. 

 The largest component of the Yrast band correspond to $K=0$. For $I=2\hbar$  this component amounts to  $98\%$,  the rest being shared by $K = \pm 1$. For $I=10\hbar$ the
 $K=0$ value is reduced to about $65 \%$ while the $K=+1$ increases up to $22\%$. For the $\beta$-band the $K$-components for $I=2\hbar$ are similar to the Yrast band and for $I=10\hbar$, $K=0$ amounts to  $45\%$, $K=-1$ amounts to  $30\%$ and $K=-2$ amounts to  $10\%$.  Finally, the $\gamma$ band has mosty $K=\pm 2$ components and small admixtures of other $K$-values which increase with spin.  
  
\begin{figure}[htbp]
\centering
\includegraphics[width=0.48\textwidth]{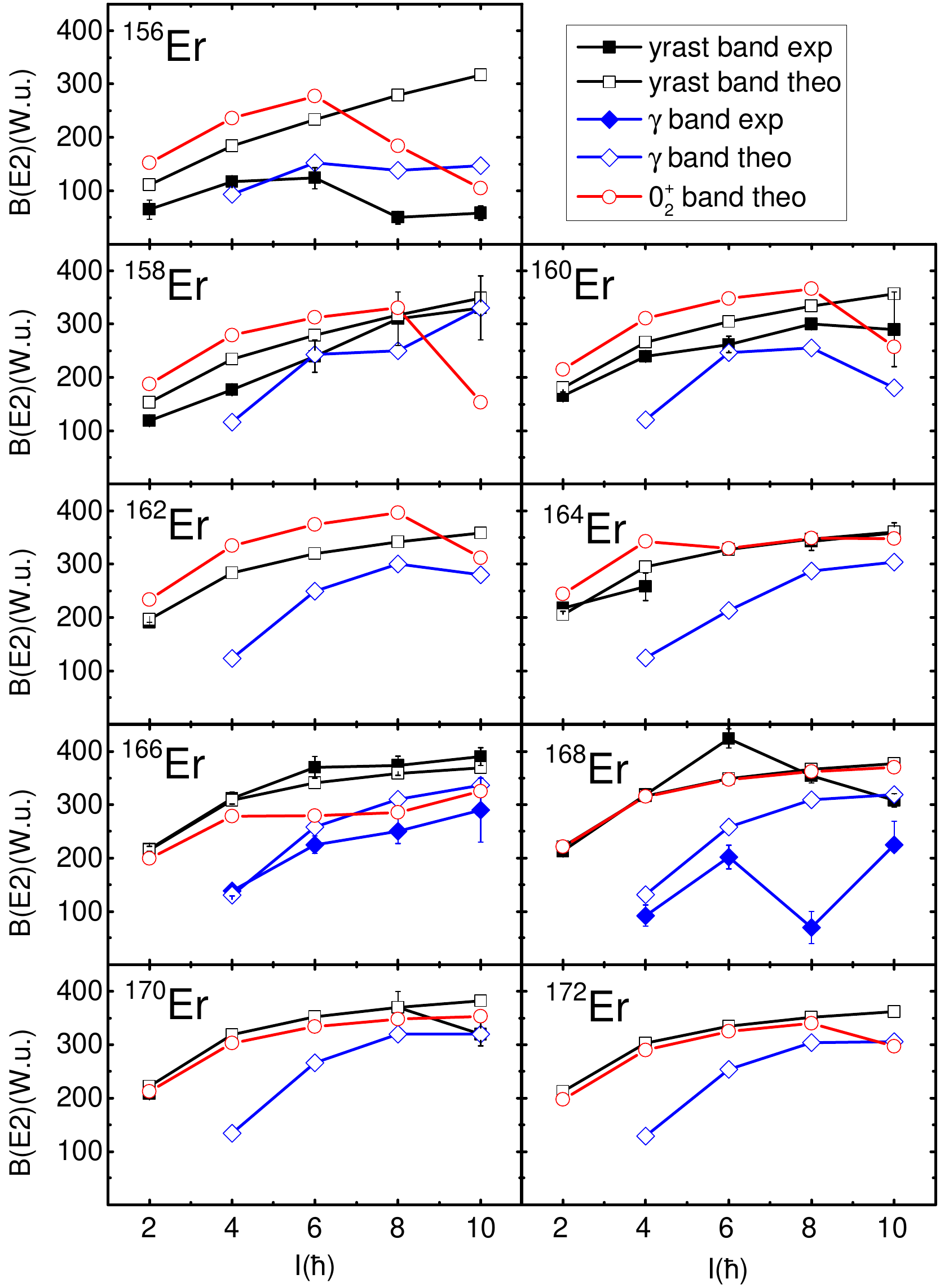}
\caption{(Color online) Calculated $B(E2,I\rightarrow I-2)$ values along the Yrast band,  the $0^{+}_{2}$ band and the $\gamma$-band of $^{156-172}$Er, compared with available data. The data are taken from \cite{ENSDF}.}
\label{ErBE2}
\end{figure}


\subsection{Transition probabilities}
Further relevant information is provided by  the intraband $E2$ transition probabilities. In Fig.~\ref{ErBE2}  we plot the calculated $B(E2,I\rightarrow I-2)$ values along the Yrast, the $\beta$ and the $\gamma$  bands  together with the available experimental information.  We first discuss the Yrast bands.  With the exception of $^{156}$Er at the largest spin values, the calculated results, black empty squares,  are in reasonable agreement with the data, black filled squares.  In particular, the transition  probability from the
 $2^{+}_{1}$  to the  $0^{+}_{1}$ state increases roughly with the neutron number, corresponding to the transition towards the well deformed region. The saturation of the experimental $B(E2)$ values at medium spins is qualitatively reproduced by the theoretical results.   

  The theoretical values for the intraband $B(E2)$ along the $0^{+}_{2}$ bands, red empty circles,  are also shown in the  Fig.~\ref{ErBE2} . For the lighter isotopes their values are larger  than the corresponding ones along the yrast bands. This can be understood by the fact that 
  the maximum of the wave function of the $\beta-$bands, see Fig.~\ref{wf_156-164_vac}-\ref{wf_156-164_all}, peak at larger deformation values than the ground states.
   In the heavier isotopes the $B(E2)$ along the $0^{+}_{2}$  are similar or smaller than along the Yrast band.  This corresponds to the above mentioned fact that for the heavier nuclei $^{166-172}$Er the wave function of the $0^{+}_{2}$ band peaks at the same deformation as the ground state due to the rigidity of the PES.  Furthermore if one takes into account that the wave functions of these isotopes contain a large amount of two-quasiparticle components, see Fig.~\ref{2qp_tot}, one understands that the $B(E2)$ along the $0^{+}_{2}$ band can be smaller than along the Yrast band. The decrease at around $I=10\hbar$ is due to the band crossing with a two-quasiparticle band.

The theoretical values of the $E2$ transition probabilities along the $\gamma-$band,  blue empty rhombi, are, in general, smaller than the corresponding values along the Yrast and the $\beta$-bands. This is also the case for the available experimental values, blue filled rhombi, compared with the analogous values along the other bands.  Theory and experiment agree reasonably well with the exception of the spin value $I=8 \hbar$ in $^{168}$Er. This discrepancy is probably due to a two-quasiparticle band crossing that we do not have in our calculations at this spin value.

\begin{figure}[htbp]
	\centering
	\includegraphics[width=0.40\textwidth]{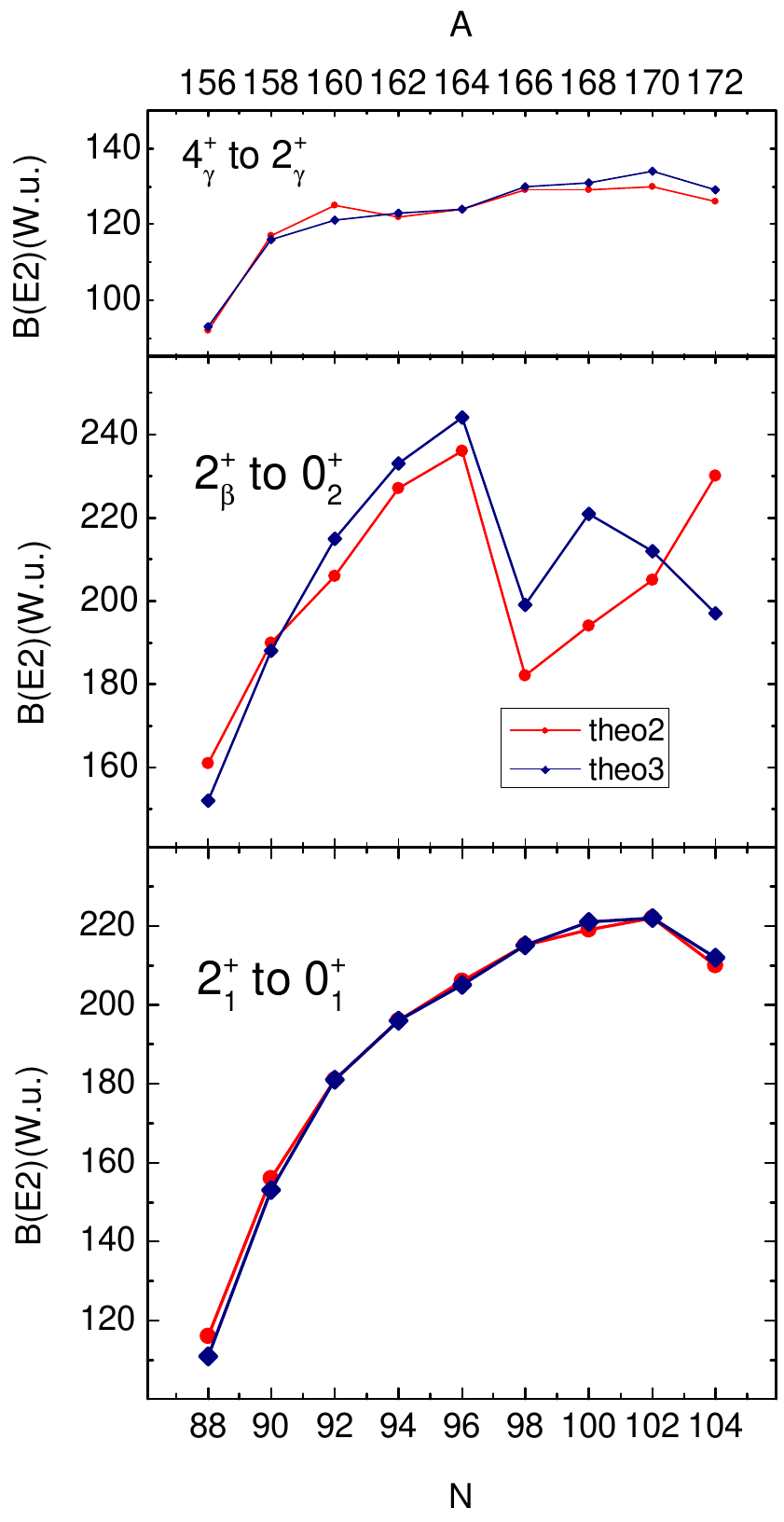}
	\caption{(Color online) Top panel: Calculated $B(E2,4^{+}_{\gamma}\rightarrow 2^{+}_{\gamma})$ for the Erbium isotopes in the "theo2" and "theo3" approaches.  Middle panel: same as top panel but for the transitions $B(E2,2^{+}_{\beta}\rightarrow 0^{+}_{2})$. Bottom panel:  same as top panel but for the transitions $B(E2,2^{+}_{1}\rightarrow 0^{+}_{1})$. }
	\label{fig2-BE2_new}
\end{figure}

In Fig.~\ref{GCMvs2QP} we have discussed the impact of the 2qp states on the excitation energies of the first excited state of the collective bands.   Now we would like to know the influence of the 2qp states on the transition probabilities. In Fig.~\ref{fig2-BE2_new} we display the indicated transition probabilities in the approaches "theo2", Eq.~(\ref{eqwf1}),  and "theo3", Eq.~(\ref{eqwf}). The results for the $\gamma$-band and the ground band are rather similar in both approaches. For the $\beta$-band transitions we find that the 2qp states provide, in general,  an increase  of about 5\%-10\%.

\begin{figure}[htbp]
\centering
\includegraphics[width=0.5\textwidth]{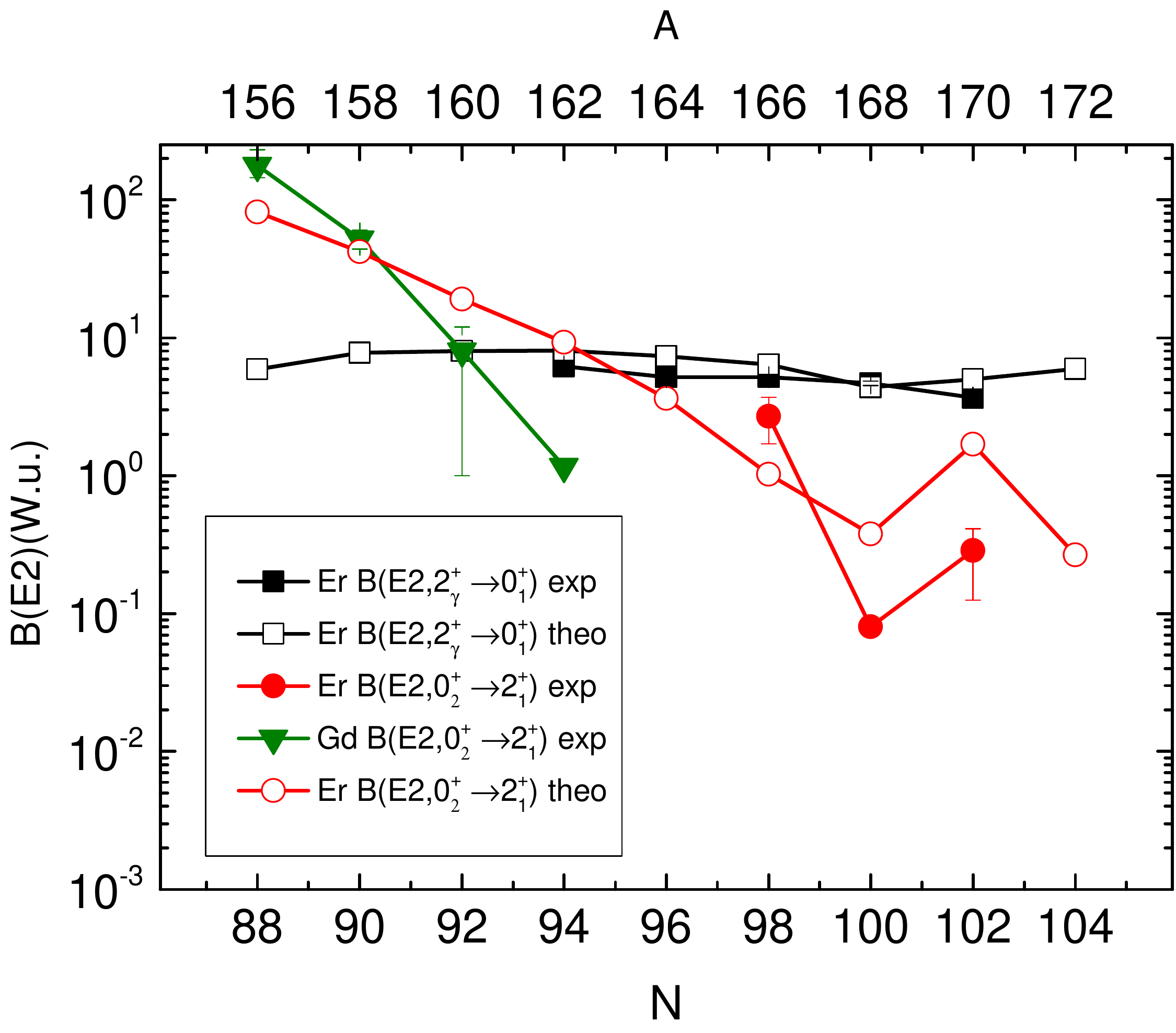}
\caption{ (Color online) Calculated $B(E2,0^{+}_{2}\rightarrow 2^{+}_{1})$ and $B(E2, 2^{+}_{\gamma}\rightarrow 0^{+}_{1})$ values for $^{156-172}$Er, compared with experimental data for Erbium and Gadolinium isotopes  with the same neutron numbers as the Er isotopes. The data are taken from \cite{ENSDF} except for  $^{166}$Er, $^{168}$Er and $^{170}$Er, which is taken from Ref.~\cite{Er166dat},  Ref.~\cite{Er168BE2} and Ref.~\cite{Er170BE2}, respectively.}
\label{ErBE2crossing}
\end{figure}

Other transitions that provide important clues are the  $E2$ transitions connecting the lowest  member of the excited bands and the ground band, namely the interband $B(E2, 0^{+}_{2}\rightarrow 2^{+}_{1})$ from the head of $\beta-$band to the $2^{+}$ state of the ground band and the 
interband $B(E2, 2^{+}_{\gamma}\rightarrow 0^{+}_{1})$ from the head of $\gamma-$band to the ground state.  
These  transition rates are related to the quadrupole collectivity of the initial state.
The calculated interband $B(E2)$ values are shown in Fig.~\ref{ErBE2crossing} together with the data measured in Erbium isotopes.
 Let us first study the decay from the $\beta-$band for the different isotopes and compare with the experimental data. For the Er isotopes these transitions probabilities have  only been measured for $^{166}$Er \cite{Er166dat}, $^{168}$Er \cite{Er168BE2} and $^{170}$Er \cite{Er170BE2}. 
Our theoretical values provide a qualitative agreement with these three values.
In order to examine the neutron number dependence of this interband $B(E2)$, we also include  in Fig.~\ref{ErBE2crossing}  the values of these transitions measured in some Gd isotones.  
 It is found that the interband $B(E2)$ values, if the $Z$ dependence is ignored, decrease with growing neutron number. This trend is reproduced by our calculations. For $^{156}$Er with $N=88$, the calculated interband $B(E2)$ is around 100 W.u., which is of the order estimated for a $\beta$-vibration \cite{Garrett} (although the estimation is made with the assumption of well deformed nucleus). This suggests that the $0^{+}_{2}$ state in this nuclei is dominated by shape vibrations. On the other hand, the calculated interband $B(E2)$ for $^{168}$Er is about 0.4 W.u., which suggests that in this case the $0^{+}_{2}$ state is dominated by two-quasiparticle excitations. These suggestions are in accordance with what was inferred from Fig.~\ref{GCMvs2QP}, see also Sect.~\ref{2qp_distri}. Note that the variation of the interband $B(E2)$ within these nuclei is as large as three orders of magnitude, which indicates that these $0^{+}$ excitations are of very different structure. This large variation range has been well reproduced by our calculation, which is another justification of our GCM+2QP model space.

We now discuss the  reduced transition probabilities from the $\gamma$ band head to the ground state $B(E2, 2^{+}_{\gamma}\rightarrow 0^{+}_{1})$. In Fig.~\ref{ErBE2crossing}  we show the experimental values, black filled squares, and the theoretical ones, black empty squares.  The agreement between theory and experiment is qualitatively good.  An intriguing  aspect is that in this case the isotopic dependence of the $B(E2, 2^{+}_{\gamma}\rightarrow 0^{+}_{1})$ is rather constant and therefore different from that of  $B(E2, 0^{+}_{2}\rightarrow 2^{+}_{1})$. We can understand the differences between the two cases cases by the following observations: For the lighter Er isotopes and at small spin values,  the intraband transitions are twice as large for the $\beta-$ as for the $\gamma-$band, see Fig.~\ref{ErBE2},  and  the wave function of the $0^{+}_2$ state peaks at larger $\beta$ deformation than the $2^{+}_{\gamma}$, see Fig.~\ref{wf_156-164_vac}. That means, for these nuclei, the state $0^{+}_2$ is much more collective than
the $2^{+}_{\gamma}$. This  explains the enhancement of the $B(E2, 0^{+}_{2}\rightarrow 2^{+}_{1})$  compared with $B(E2, 2^{+}_{\gamma}\rightarrow 0^{+}_{1})$ for the lighter isotopes.    For the heavier isotopes, the largest content of 2qp in the $0^{+}_2$ state as compared with the $2^{+}_{\gamma}$ state, see Fig.~\ref{Er2qpweight}, will explain the drop 
in the $B(E2, 0^{+}_{2}\rightarrow 2^{+}_{1})$.

\section{Summary and conclusions}\label{sect4}

In this work the ansatz of the generator coordinate method has been generalised, for the first time, as to include 
particle number and angular momentum symmetry conserving fluctuations in the $(\beta,\gamma)$ plane and their symmetry conserving two-quasiparticle excitations.  This ansatz allows us to investigate microscopically the role played by the collective and the single-particle degrees of freedom in the most relevant states of the atomic nucleus, namely, the Yrast$-$,  $\beta-$ and
 $\gamma-$bands.  This theory has been applied to the $N=88-104$ Erbium isotopes. This region includes very soft nuclei, like $^{156-158}$Er, strong deformed ones,  like $^{164-172}$Er, and the transition nuclei in between.  
In the calculations the separable pairing plus quadrupole Hamiltonian has been used.  

 The calculated spectra of the Yrast$-$ and the $\beta-$bands are in good agreement with the experimental data. The  theoretical $\gamma-$bands, in line with other approaches \cite{Tera}-\cite{Bohr_5DM}, are somewhat higher than the experimental data, we obtain however a good agreement for the moments of inertia. 
 
 Concerning the transition probabilities,  the available experimental intraband $E2$ reduced transition probabilities along the bands  are reasonably well reproduced (except for $I=8 \hbar-10 \hbar$ in $^{156}$Er along the Yrast band and for $I=8 \hbar$ in $^{168}$Er for the $\gamma-$band). The transition from soft to rigid deformation with increasing neutron number is also well reproduced by our calculations.
 The interband   $B(E2,0^{+}_{2}\rightarrow 2^{+}_{1})$ and $B(E2, 2^{+}_{\gamma}\rightarrow 0^{+}_{1})$ display a disparate behaviour. While the $B(E2,0^{+}_{2}\rightarrow 2^{+}_{1})$ values show variations of up to three orders of magnitude in the studied isotopes and Gd isotones, the   $B(E2, 2^{+}_{\gamma}\rightarrow 0^{+}_{1})$ values are more or less constant in the same interval. These large variations are properly described in our approach.
 
 A detailed look at the collective wave functions and the separation of the contributions of the HFB vacua and the two-quasiparticle states allows us to analyse the characteristic of the band heads. The ground states $(0^{+}_{1})$, in general,  have a small component of two-quasiparticle states in their wave functions for all isotopes. In the lighter isotopes, specially $^{156-158}$Er, the wave functions display broad distributions in the $(\beta,\gamma)$ plane which become more concentrated  as the neutron number increases to the well deformed heavier isotopes.  The band head of the $\beta-$band has a different character. The relevance of the quasiparticle states is  small for the soft nuclei and very large for the rigid ones.  This change in the composition of the wave function is able to explain the three orders of magnitude variation found in the $B(E2,0^{+}_{2}\rightarrow 2^{+}_{1})$.
 The genuine $\beta$ vibration is founded to be strongly coupled to the triaxiality and the two-quasiparticle states. The characteristic node of the one-dimensional degree of freedom (the deformation parameter $\beta$) turns out to be a valley in the $(\beta,\gamma)$ plane.  Concerning  the relevance of the quasiparticle states in the $\gamma$ band head of the Er isotopes, in the lighter, softer isotopes the relevance is small.  In the heavier, rigid ones it increases steadily but it never becomes  as large as in the $\beta$-band. For the soft Er isotopes the  $\beta-$band is more collective than the $\gamma-$band while for the heavier isotopes they are comparable.
 
 In conclusion, the presented approach provides an overall good agreement with the experimental data of nine erbium isotopes ranging from very soft to very rigid shapes. Furthermore, it allows a comprehensive understanding of the microscopic structure of the most relevant states of the atomic nucleus.  The only issue not quite understood is the
 reason why in the present and others approaches the predicted excitation energy of the $\gamma$ band head is systematically higher than what is  experimentally observed.

\begin{acknowledgments}
 This work was supported by the Spanish Ministerio de Econom\'ia y Competitividad under contracts FPA2011-29854-C04-04 and FPA2014-57196-C5-2-P.

\end{acknowledgments}

\end{document}